# Density-wave phases, anisotropic transport, and Planckian dissipation in single crystals of the superconductor $La_3Ni_2O_7$

Zhehong Liu[1,2†*], Masamichi Nakajima[1†], Markus Kriener[1], Shunsuke Kitou[3], Xiaowei Lyu[1], Chieko Terakura[1], Kosuke Karube[1], Ka Man Yip[4], Sorin Lazar[4], Nobuto Nakanishi[5], Keiko Shimada[1], Akiko Kikkawa[1], Yukako Fujishiro[1,6,7,8], Xiuzhen Yu[1], Taka-hisa Arima[1,3], Yoshinori Tokura[1,6,9*], Yasujiro Taguchi[1,2*]

[1] *RIKEN Center for Emergent Matter Science (CEMS), Wako 351-0198, Japan*

[2] *RIKEN Baton Zone Program, TRIP Headquarters, Wako 351-0198, Japan*

[3] *Department of Advanced Materials Science, The University of Tokyo, Kashiwa 277-8561, Japan*

[4] *Materials and Structural Analysis Division, Thermo Fisher Scientific, Achtseweg Noord 5, Eindhoven, Netherlands*

[5] *Materials and Structural Analysis Division, Thermo Fisher Scientific, FEI Company Japan Ltd. Tokyo140-0002, Japan*

[6] *RIKEN Pioneering Research Institute (PRI), Wako 351-0198, Japan*

[7] *Department of Applied Physics, The University of Tokyo, Tokyo 113-8656, Japan*

[8] *Institute of Engineering Innovation, The University of Tokyo, Tokyo 113-8656, Japan*

[9] *Tokyo College, The University of Tokyo, Tokyo 113-8654, Japan*

*Corresponding authors: zhehong.liu@riken.jp; tokura@riken.jp; y-taguchi@riken.jp

†These authors contributed equally

## Abstract

Pressure-induced superconductivity in bilayer nickelates provides a platform for investigating intertwined roles of charge/spin orders and electric transport in unconventional superconductivity. However, important quantitative information on the transport, such as the absolute value of the resistivity, the anisotropy, and the scattering rate of carriers, remains insufficient due to the lack of accurate measurements using large single crystals. Here we establish a high-precision pressure–temperature phase diagram of high-quality $La_3Ni_2O_7$ single crystals, by measuring the in-plane and out-of-plane resistivities. We resolve two distinct anomalies associated with density-wave formation with contrasting pressure dependences. The pressure-induced structural transition enhances not only the resistivity values for both directions, but also its anisotropy at low temperatures, demonstrating a pronounced effect of density-wave order on the charge dynamics. Superconductivity with zero-resistance emerges near the boundary where the density-wave phases are fully suppressed, and above $T_c$, the resistivity exhibits a temperature-linear dependence over a wide temperature range while the scattering rate falls within a regime of the Planckian limit. Our results show that pressure dramatically changes the anisotropic charge transport via modifying density-wave orders, and eventually produces a pronounced strange-metal state with strong scatterings, from which superconductivity develops. This establishes robust density-wave correlations and Planckian dissipation as remarkable features of $La_3Ni_2O_7$.

## Introduction

The discovery of pressure ($P$)-induced superconductivity in the Ruddlesden-Popper series of nickelates represented by bilayer $La_3Ni_2O_7$ has provided a new arena for exploring unconventional superconductivity in correlated transition-metal oxides [1-6] because the superconducting transition temperature ($T_c$) approaches that of several unconventional high-$T_c$ superconductors [7-11]. The $NiO_2$ bilayer structure and multiorbital electronic state, with strong magnetic and electron-lattice interactions, place $La_3Ni_2O_7$ at the intersection of important topics in correlated-electron physics, such as density-wave (DW) order, strange-metal transport and superconductivity. Elucidating how these phenomena evolve together as a function of pressure ($P$) is essentially important for identifying the microscopic mechanisms that lead to superconductivity [1,2,4,12-14].

In several unconventional superconductors, such as cuprates, charge-density-wave (CDW) and spin-density-wave (SDW) correlations largely affect the electronic structure, compete with superconductivity and, in some cases, may serve as the fluctuations that mediate pairing [15-20]. Transport studies of $La_3Ni_2O_7$ have reported $P$-dependent anomalies that are attributed to DW order, structural transformation, and the onset of superconductivity [1,2,4]. However, the reported $P$-temperature ($T$) phase diagrams remain incomplete, partly due to limitations in single-crystal quality and reproducibility [1,2,21-23].

Two anomalies related to DW order have been reported [24-26]. The higher-$T$ anomaly ($T_2$) has been associated with SDW order, while the lower-$T$ one ($T_1$) with an additional DW-like state whose microscopic character remains unclear. Previous transport phase diagrams generally suggested that $T_1$ decreases monotonically with increasing $P$ and disappears below approximately 10 GPa [1,2,4,24]. By contrast, the $P$-evolution of $T_2$ remained unclear [24,25] until recently [27]. To establish clear correlations of such DW orders with the electric transport characteristic and superconductivity is of prime importance.

Pressure also drives an orthorhombic-to-tetragonal structural transition before superconductivity emerges, which modifies the one-electron energy levels, orbital hybridization and interlayer coupling, *etc.* [1,4], possibly leading to changes of the anisotropic electronic properties. Another important issue is the scattering mechanism in the metallic normal-state above $T_c$ under high-$P$. A $T$-linear resistivity over an extended $T$-range is observed as defining feature in various unconventional superconductors [10,28]. In many cases, the inferred inelastic scattering rate is comparable to the Planckian scale, $\hbar/\tau \sim k_B T$, suggesting that charge relaxation is governed by a nearly universal timescale dictated by temperature [28-33]. A quantitative determination of the scattering rate in nickelates is needed because it may reveal the nature of the strongly correlated normal state with a possible link to quantum critical fluctuations, such as robust CDW/SDW fluctuations, in the perspective of other unconventional superconductors.

Here we establish a high-precision $P$–$T$ transport phase diagram of $La_3Ni_2O_7$ single crystals through systematic high-$P$ measurements of the in-plane resistivity $\rho_{ab}$ and the out-of-plane resistivity $\rho_c$. The high quality of our single crystal enables us to resolve DW-related anomalies over a broader *P*-range. We find that $T_1$ does not vanish below 10 GPa [1,2,4], and can be tracked across the structural transition around 8-10 GPa [34] with an abrupt increase near the boundary. In parallel, both $\rho_{ab}$ and $\rho_c$, as well as the resistivity anisotropy $\rho_c/\rho_{ab}$, increase upon traversing this transition at low *T*s. In addition, $T_2$ increases monotonically with pressure, revealing a robust SDW order that persists deep in the tetragonal phase [24,27].

At higher *P*s where DW instabilities are eventually suppressed, normal-state resistivity displays a prominent *T*-linear dependence in a wide *T*-range from the onset temperature of superconductivity ($T_c^{onset}$) up to 300 K. The extracted values of the scattering rate are quantitatively consistent with Planckian dissipation, showing that $La_3Ni_2O_7$ belongs to a broad class of strange metals.

## Oxygen-stoichiometric and stacking-faults-free bilayer $La_3Ni_2O_7$ single crystals.

Oxygen non-stoichiometry and intergrowths of multiple Ruddlesden-Popper phases have frequently been encountered in $La_3Ni_2O_7$ [1,5,6,23,35,36], obscuring the intrinsic electronic behavior of the bilayer system. We overcome this issue and obtain high-quality, large-sized bilayer $La_3Ni_2O_7$ single crystals. As shown in **Fig. 1a**, $La_3Ni_2O_7$ crystallizes in an orthorhombic structure at ambient *P*. Recently, a checkerboard-type charge order with two distinct valence states of Ni on the $NiO_2$ planes at room temperature has been reported [37], and the space group of the structure is *Am*2*m.* Synchrotron X-ray diffraction and subsequent structure refinement (**Supplementary Figs. S4** and **S5**) unambiguously confirm the *Am*2*m* structure for our single-crystal as well, and reveal evidence for the absence of oxygen vacancies within experimental accuracy (see detailed structural information in **Supplementary Table S1-S3**).

Polarized optical microscopy reveals flat, well-developed *ab*-plane cleavage surfaces with homogeneous contrast (**Fig. 1b**). Typical crystals reach lateral dimensions of ~220 μm. The Laue diffraction pattern (**Fig. 1c**) exhibits sharp and virtually 4-fold symmetric spots, attesting to an *ab*-face with high crystallinity, consistent with the X-ray diffraction results showing only sharp (00*l*) reflections (**Supplementary Fig. S3**). The *c*-axis lattice parameter of the plate-like specimen is determined as 20.5413(13) Å, in good agreement with our refined structure (**Supplementary Table S1**) for a 50-μm crystal as well as previous reports [1,2,4,37].

Imaging by transmission electron microscopy (TEM) reveals a uniformly ordered $NiO_2$ bilayer structure throughout the investigated region (**Fig. 1d**). Extensive examinations over multiple areas and large fields of view find no evidence of intergrowths of monolayer, trilayer, or the so-called 1313 and 1212 phases, which have commonly been reported in earlier studies [5,6,23,35]. Energy-dispersive X-ray spectroscopy (EDS) mapping over the entire crystal, together with local EDS measurements (**Supplementary Fig. S1**) reveals a homogeneous element distribution. Quantitative EDS analyses at multiple locations (**Supplementary Data Fig. S2**) yield a nearly constant La : Ni ratio close to the nominal 3:2 stoichiometry. These results confirm the compositional homogeneity of crystals of the bilayer $La_3Ni_2O_7$ phase.

## Transport signature of density-wave phases and high-$T_c$ superconductivity

We next discuss the *P*-evolution of in-plane (*ab*-plane) transport of $La_3Ni_2O_7$ (sample 2#-a). Representative data for selected *P* values are plotted as a function of *T* on a logarithmic scale in **Fig. 1e**.

As *P* increases, $\rho_{ab}$ at 300 K deceases very rapidly from the order of $10^{-2}$ Ω cm with a metallic *T*-dependence at low pressures (0-8 GPa) to the order of $10^{-3}$ Ω cm with a semiconducting-like *T*-dependence at intermediate pressures (12-16 GPa), accompanied by the onset of superconductivity at low *T*s (**Fig. 1e**). Upon further increasing *P* to 20 GPa, $\rho_{ab}$ at 300 K is as low as $10^{-4}$ Ω cm with a fully recovered metallic *T*-dependence, while superconductivity is observed with an onset *T* of 68 K.

To show the detailed behaviour, we plot the resistivity data for all *P*-values on a linear scale in **Figs. 2a-c**. As *P* is increased at the low-*P* range (0-8 GPa), a rapid change of the resistivity is observed around 300 K while the residual resistivity remains almost constant (**Fig. 2a**). The rapid decrease at 300 K is attributed to a *P*-induced collapse of the charge order [34]. On entering the high-*P* regime, a semiconducting-like upturn develops at relatively high *T*s (250-300 K) and also at low $T$s below 150 K, while a superconducting transition emerges at lower $T$ (**Fig. 2b**). This qualitative change occurs near 8-10 GPa, coincident with the structural transition [34]. Upon further compression to 12-16 GPa (**Figs. 2b,c**), the semiconducting-like $T$-dependence extends over a broad $T$-range of 70-300 K. Concurrently, the superconducting transition becomes progressively more pronounced, accompanied by a systematic increase in $T_c^{onset}$. Even at 16 GPa where superconductivity starts to set in near 63 K, the normal-state resistivity still exhibits a pronounced low-$T$ upturn towards $T_c^{onset}$. Upon further increasing the pressure to 18-20 GPa, $\rho_{ab}$ eventually shows a totally metallic state with $T$-linear dependence over 80-300 K (**Fig. 2c**), signaling the emergence of a strange-metal-like state. The pronounced evolution of the transport characteristics including robust semiconducting-$T$ behaviour highlight a substantial and non-trivial $P$-driven modification of the electronic state and DW orders.

The $P$-dependence of $\rho_{ab}$ further resolves the distinct roles of the DW orders. At 290 K, $\rho_{ab}$ decreases progressively and sharply drops between 12 and 14 GPa (**Fig. 2d**). By contrast, at 10 and 70 K, $\rho_{ab}$ initially decreases slightly below 6 GPa, but rather increases around 8-10 GPa in the vicinity of the structural transition.

Analysis of the resistivity and its $T$-derivative at different $P$s (**Extended Data Fig. 1**) enables us to identify and track the characteristic $T$s. Following a previous report [24], we assign two anomalies in resistivity data at ambient pressure to DW ($T_1$) and the SDW ($T_2$) order. The resulting $P$-$T$ phase diagram reveals that the two DW anomalies exhibit markedly different $P$-dependences (**Fig. 2e**). The DW anomaly $T_1$ is initially suppressed upon compression but increases sharply around 10 GPa in the tetragonal phase, similar to refs. [12,13]. The higher-$T$ SDW anomaly $T_2$ instead increases monotonically with increasing $P$. The $T_1$ and $T_2$ remain discernible in the $P$-regime below ~18 GPa where at low-$T$ a semiconducting-like $T$-dependence is observed. Both anomalies disappear when the system recovers a fully metallic, $T$-linear resistivity at higher $P$s. In addition to $T_1$ and $T_2$, three temperatures representing the resistivity minimum, maximum, and sharp resistivity upturn at high-$T$s are plotted as characteristic $T$s on the phase diagram. Furthermore, tiny anomalies are observed at lower $T$ than $T_1$ (see **Extended Data Fig. 1**).

Superconductivity appears exclusively in the tetragonal phase, with $T_c^{onset}$ and the zero-resistivity temperature ($T_c^{zero}$) reaching approximately 68 K and 55 K at 20 GPa, respectively. Notably, the zero-resistance state emerges on the verge of the disappearance of $T_1$, $T_2$, and the sharp upturn of the resistivity.

## Pressure-dependent anisotropic transport

To elucidate the $P$-driven change of the electronic state, we measure both the in-plane $\rho_{ab}$ and out-of-plane $\rho_c$ resistivities for another crystal 2#-e by using the Montgomery method [38], and determine the transport anisotropy, $\rho_c/\rho_{ab}$.

The in-plane transport of crystal 2#-e reproduces the key features observed in crystal 2#-a (**Fig. 3a,b**). At low pressures (0-8 GPa), $\rho_{ab}$ exhibits a bad-metallic $T$-dependence and decreases progressively at high $T$s with increasing $P$. At 10 GPa, however, $\rho_{ab}$ remains metallic at high-$T$s while it shows a semiconducting-like $T$-dependence at low-$T$s. This low-$T$ upturn, which is also observed in crystals 2#-a and 2#-d (**Extended Data Figs. 1,5,6**), is assigned to the low-$T$ DW order. Upon further increasing $P$ to 12-20 GPa, the low-$T$ resistivity continues to exhibit a semiconducting-like $T$-dependence below the resistivity minimum while superconductivity emerges and its onset $T$ increases. At 18-20 GPa, a semiconducting-like $T$-dependence develops at high $T$s and such a $T$-range extends towards lower $T_S$ with increasing $P$, reminiscent of the behaviour of crystal 2#-a at 12-14 GPa (**Fig. 2b**).

$\rho_c$ shares the same $P$-dependence of $\rho_{ab}$ at low $T$s, but differs markedly at high $T$s (**Figs. 3c,d**). At 200 K, $\rho_c$ decreases from 0 to 8 GPa, and then increases continuously from 8 GPa to 20 GPa. The initial strong reduction of $\rho_c$ may reflect the collapse of charge order. The increase of $\rho_c$ at around 200 K and in the high-$P$ region may indicate a $P$-induced non-trivial change of electronic hybridization.

This modification is more clearly captured by $\rho_c/\rho_{ab}$ (**Figs. 3e,f**). Between 0 and 6 GPa, the anisotropy generally increases upon cooling from 10-13 to 20-25, and overall remains at similar values with increasing pressure.

Near 8-10 GPa, the anisotropy exhibits a remarkable change. It rises sharply at low temperatures below about 150 K upon changing from 8 GPa to 10 GPa, coincident with the structural transition [34]. The increased Ni–O–Ni bond angle (= 180°) in the tetragonal phase should enhance the Ni-3*d* – O-2*p* overlap for both in-plane and out-of-plane directions. The observed large increase in anisotropy only at low-*T*s instead points to the importance of DW orders as the dominant origin of the large modification of the transport anisotropy.

At higher *P*s, the anisotropy becomes strongly *T*-dependent. Between 12 GPa and 20 GPa, it decreases and increases with pressure below ~100 K and above ~150 K, respectively. Representative isotherms further highlight this dichotomy (**Fig. 4a**). At 290 K, $\rho_c/\rho_{ab}$ increases gradually and reaches nearly 90 at 20 GPa. At 10 and 70 K, it remains nearly constant from ambient pressure up to ~8 GPa, and sharply increases to approximately 100 upon entering the tetragonal phase at 10 GPa. Above 12 GPa, the anisotropy decreases from 12 GPa to 20 GPa although it is affected by superconductivity at 10 K. This trend is consistent with the suppression of the low-*T* DW correlations ($T_1$) that may progressively enhance the interlayer hopping. More detailed *P*-dependences of $\rho_{ab}$, $\rho_c$, and $\rho_c/\rho_{ab}$ at selected temperatures, and the corresponding *P*-*T* phase diagrams are presented in **Extended Data Figs. 4a-f**. Very similar temperature-dependent transport behaviors are also observed in another single crystal (2#-d), as presented in **Extended Data Figs. 5-8**.

These anisotropic transport data show that the *P*-induced structural change modifies the DW orders at low-*T*s, which in turn affects the interlayer transport of the conduction electrons. One possible origin would be a *P*-induced change of the stacking of the DW orders along the *c*-axis. A recent neutron scattering study for $La_3Ni_2O_7$ [39] reported coexistence of two different *q*-vectors with the same in-plane pattern and a distinct stacking along the *c*-axis. A possible *P*-induced change of the *c*-axis stacking may account for the observed large change of the transport anisotropy at low-*T*s. Alternatively, another resonant X-ray scattering study [40] indicates a short correlation length of the SDW order along the *c*-axis, which may prevent coherent electron transport along the interlayer direction, leading to confined two-dimensional superconductivity. Consequently, a sharp decrease in $\rho_{ab}$ is observed at $T_c^{onset}$, although the resistivity remains finite. This bears some analogy to the stripe-ordered cuprates with dynamically decoupled $CuO_2$ layers [41]. The electrical resistivity anisotropy of $La_3Ni_2O_7$ is on the order of several tens, comparable to that of optimally doped $YBa_2Cu_3O_7$ (YBCO), but significantly smaller than those of more anisotropic cuprates, including $La_{1.875}Ba_{0.125}CuO_4$ (LBCO, 214; $10^3$~$10^5$) and $Bi_2Sr_2CuO_6$ (Bi-2201; > $10^4$) [41,42].

**Reproducibility and sample-dependent pressure response**

To check the reproducibility of our results, we compare the high-*P* transport data of single crystals (2#-a, 2#-d and 2#-e) and polycrystalline material [27]. As plotted in **Fig. 4b**, the general features of the phase diagrams are highly reproducible across the samples. In particular, the SDW order at $T_2$ increases monotonically with *P* in both the single crystals and the polycrystalline sample. The increase of $T_2$ is interpreted in terms of an enhanced exchange interaction, similarly to those observed in correlated materials on the verge of a bandwidth-controlled Mott transition, such as $R$NiO$_3$ (*R* = Pr, Nd, Sm, Eu) and $NiS_{2-x}Se_x$ [43-45]. The DW order at $T_1$ also follows the same overall evolution in the three single-crystals. The polycrystalline sample follows this low-*P* evolution, whereas the $T_1$ anomaly becomes unclear above ~ 5 GPa. These observations show that the *P*-driven variation of the long-range DW orders is common while the sensitivity to the low-*T* anomaly depends on the sample form (single-crystalline vs. poly-crystalline). Meanwhile, as shown in **Extended Data Figs. 2,5b**,**9b**, and **Supplementary Table S4**, the overall resistivity behavior, resistivity values at corresponding pressures, and the characteristic temperatures $T_c^{onset}$ and $T_c^{zero}$ ($T_c^{mid}$) at 20 GPa are highly consistent among the three single crystals.

While the overall phase diagram is reproducible, a slight sample-dependent response is observed. The *P*-evolution of $T_c^{onset}$ in crystal 2#-a closely follows that in crystal 2#-d and in the polycrystalline sample [27] while $T_c^{onset}$ shifts to higher *P*s in crystal 2#-e. Such offsets are also discernible in the normal-state transport. As shown in **Figs. 4d,e**, the resistivity behaviour of crystal 2#-e at 18 GPa and 20 GPa closely resembles that of crystal 2#-a at 12 GPa and 14 GPa, respectively, in terms of their magnitudes, *T*-dependences, and superconducting onsets. The corresponding high-*T* crossover as indicated by arrows shifts similarly. Both the crystals exhibit a *P*-induced steady suppression of the resistivity values while they suffer from robust semiconducting-like upturns at low- and high-$T$s. Crystal 2#-a eventually exhibits $T$-linear resistivity and superconductivity upon further increasing $P$ to 18 GPa and 20 GPa.

The slight sample-dependence of the $P$-offsets may likely reflect tiny differences in local strain, carrier density, or residual microscopic disorder. Nevertheless, the reproducibility of the overall phase diagram establishes the intrinsic nature of the $P$-driven electronic modification in $La_3Ni_2O_7$.

## Planckian dissipation

As shown in **Fig. 4f**, the metallic state at 18 GPa and 20 GPa displays the $T$-linear $\rho_{ab}$ over a wide $T$-range from 80 K to 300 K with almost zero residual resistance. Notably, the $\rho_{ab}$ value just above $T_c$ is as low as ~13 μΩ cm, highlighting the minimal disorder in the single-crystalline compound $La_3Ni_2O_7$ sample. Following Ref. [28], we determine the transport scattering rate from the resistivity slope, $d\rho/dT$ = 0.28 μΩ cm/K (**Fig. 4f**), using the Drude relation $1/\tau = ne^2\rho/m^*$ and the values of $n$ and $m^*$ obtained by an angle-resolved photoemission spectroscopy (ARPES) measurement [46]. We determine $\alpha = 1.6 \pm 0.4$ for 20 GPa data (see Supplementary for the details), where $\alpha$ is a dimensionless parameter that appears in the formula of the Planckian scattering rate $\frac{1}{\tau} = \alpha\frac{k_B T}{\hbar}$ [28].

Figure 4c compares the relationship between the normalized resistivity slope, $ne^2(d\rho/dT)/(k_B k_F)$, and the Fermi velocity $v_F$ for $La_3Ni_2O_7$ and a broad range of correlated metals, which are reproduced from Ref. [28]. The data point falls close to the $\alpha = 1$ line, placing $La_3Ni_2O_7$ in the same strong-scattering regime as well-established strange metals, possibly in the vicinity of a quantum critical point.

The Planckian dissipation indicates the largest allowed scattering rate from various thermal excitations [33], therefore our result clearly demonstrates that charge carriers still suffer from strong scattering even at high $P$s where the DW orders are fully suppressed. Superconductivity with zero-resistance state emerges out of this $T$-linear resistivity state with Planckian dissipation.

Meanwhile, a similar *T*-linear resistivity has also been reported for polycrystalline $La_3Ni_2O_7$ (Ref. 27), although the absolute value of the slope differs from that observed in our single crystals, further supporting the intrinsic nature of the strange-metal-like transport behaviour.

## Conclusion

We establish a pressure-temperature phase diagram of high-quality bilayer $La_3Ni_2O_7$ single crystals by high-pressure transport measurements. Synchrotron X-ray diffraction and transmission-electron-microscopy experiments confirm an oxygen-stoichiometric, homogeneous orthorhombic *Am2m* structure with no detectable stacking faults or other Ruddlesden–Popper intergrowths. A checker-board type charge order is observed at ambient temperature and ambient pressure.

Pressure suppresses the charge order rapidly, and drives a pronounced modification of the electronic state near the orthorhombic-to-tetragonal structural transition around 10 GPa. The low-temperature density-wave transition temperature $T_1$ is abruptly enhanced around this boundary, whereas the higher-temperature anomaly at $T_2$ associated with spin-density-wave order increases rather continuously with pressure. Pressure- and temperature-dependent changes in $\rho_{ab}$, $\rho_c$, and $\rho_c/\rho_{ab}$ reveal that the effect of the structural transition on the transport anisotropy at 290 K is rather weak, and the variation in density-wave order(s) rather gives a strong impact on the anisotropy at low temperatures. Superconductivity emerges exclusively in the tetragonal phase, following this dramatic change of the resistivity and its anisotropy.

At higher pressures, $La_3Ni_2O_7$ exhibits a temperature-linear resistivity over a wide temperature range with a scattering rate consistent with Planckian dissipation, clearly revealing the strange-metal-like nature of its normal state. The overall density-wave transition hierarchy, remarkable change of the transport properties, and the superconducting phase are common across single crystals and polycrystalline samples, despite tiny sample-dependent shifts in pressure values and transition temperatures. Our present results reveal that superconductivity in $La_3Ni_2O_7$ emerges from a highly correlated electronic state with Planckian dissipation when the robust density-wave orders are fully suppressed.

## Methods

### Single-crystal growth

All crystals were grown at high pressure around 3 GPa. High-purity starting material $La_2O_3$ (99.99%) was heated at 1073 K for 10 h before use. $La_2O_3$ and NiO (99.99%) powders with molar ratio 3:4 were weighed, mixed, and ground for 3 h. After thoroughly griding, the mixture were mixed with flux and an oxygen agent and then pressed into a platinum capsule and treated at 3-4 GPa with a cubic anvil-type high-pressure apparatus, then heated at 1373 K for one hour, and slowly cooled to 1073 K. After the heat treatments, the samples were quenched to room temperature and the pressure was released slowly.

### Single crystal XRD measurement

X-ray diffraction (XRD) measurements were conducted on an in-house x-ray diffractometer (Rigaku) with copper as target source.

### Laue measurement

Laue measurements were conducted at room temperature using an m-Laue diffractometer (Pulstec Industrial Co., Ltd.) with tungsten as the target source.

### SEM measurements and EDS mapping

The chemical composition of the $La_3Ni_2O_7$ single crystal was determined using a scanning electron microscope equipped with an energy-dispersive X-ray spectrometer (SEM–EDS; JEOL Ltd. and Bruker). EDS measurements were performed at up to ten different surface regions of the same crystal to assess compositional homogeneity.

### STEM measurements and EDS mapping

The (110) $La_3Ni_2O_7$ thin specimen was extracted from a bulk single crystal using an in situ micromanipulator (Easy Lift) inside a dual-beam system equipped with a Ga ion beam (FIB) and an electron beam (Helios 5 UX, Thermo Fisher Scientific), and was subsequently thinned to a thickness below 50 nm using $Ga^+$ FIB. High-angle annular dark-field scanning transmission electron microscopy (HAADF-STEM) imaging was then performed using transmission electron microscopes (Talos F200X G2 and Iliad Ultra, Thermo Fisher Scientific) to elucidate the bilayer structure of $La_3Ni_2O_7$. Simultaneously, EDS measurements were acquired from the same sample region used for HAADF-STEM and integrated differential-phase-contrast (iDPC) observations to verify the elemental composition. EDS data is taken with probe current 80 pA, probe semi-convergence angle 21.4 mrad, 50 us dwell time per pixel, 128×128 scan size, 26.1 pm pixel size, field of view 3.34 nm, total 292 frames with drift-correction enabled. The navigation image on HAADF is taken at 36 mm camera length, while HAADF/STEM-iDPC data is taken with probe current 80 pA, probe semi-convergence angle 21.4mrad, 10us dwell time per pixel, 1024×1024 scan size, 13.04 pm pixel size, field of view 13.3nm and 185mm camera length. The combined STEM-iDPC, HAADF-STEM, and EDS analyses confirmed the compositional homogeneity of the as-grown crystals.

**Single-crystal structure determination**

The crystal structure of $La_3Ni_2O_7$ at 300 K under ambient pressure was investigated by single-crystal X-ray diffraction at the BL02B1 beamline [47] of SPring-8, Japan. A single crystal with dimensions of 50 × 50 × 20 $\mu m^3$ was used. Diffraction patterns were recorded using a two-dimensional CdTe PILATUS detector with a dynamic range of $\sim 10^6$. Data were collected using the fine-slice method with an oscillation step of $\Delta\omega = 0.1°$. The intensities of Bragg reflections were extracted using CrysAlisPro [48]. Equivalent reflections were averaged, and structural parameters were refined using Jana2006 [49].

**Transport measurements**

Near rectangular single crystals were selected from the same batch with dimensions of 420 × 390 × 50 $\mu m^3$ (2#-a), 108 × 106 × 29 $\mu m^3$ (2#-d) and 260 × 134× 63 $\mu m^3$ (2#-e) for high-pressure resistivity measurements. Four electrical contacts were prepared using EPO-TEK H20E silver epoxy for crystal 2#-a and nanoparticle silver paste (Tanaka Precious Metal Technologies Co., Ltd.) for crystal 2#-d and 2#-e. The epoxy was cured in two steps. First, samples with freshly applied contacts were placed on a hot plate in air at 120-130°C for 15-20 min to pre-cure the epoxy. Subsequently, the samples were transferred to a mini-lamp annealer (Mila 5050Z, Advance Riko) and heated at 400°C for 10 min under flowing $O_2$. The silver paste was directly treated by a mini-lamp annealer (Mila 5050Z, Advance Riko) and heated at 400°C for 10 min under flowing $O_2$ directly. Which yielded typical two-terminal contact resistances below 10 Ω.

The samples were first characterized at ambient pressure using a Physical Property Measurement System (PPMS, Quantum Design) and subsequently measured under pressure in a custom-built cubic-anvil press. Daphne 7575 and Daphne 7676 oils served as pressure-transmitting media to maintain quasi-hydrostatic conditions during the high-pressure resistivity measurements. Resistivity measurements under pressure were performed upon warming from approximately 4.5 to 295 K. The cubic-anvil press provides a highly hydrostatic pressure environment.

For the anisotropic resistivity measurements and calculations, we used the Montgomery method [38] with a special electrode configuration (as shown in the inset of **Fig. 3a** and **Extended Data Fig. 5**). The pressure-induced lattice contraction reported in ref. 34 was taken into account in the resistivity calculations based on the Montgomery method [38].

**Acknowledgements**

This work was supported by JST CREST (Grant No. JPMJCR20T1), and the RIKEN TRIP initiative (Many-body Electron Systems, and Advanced General Intelligence for Science Program), and Sumitomo Chemical, and JST PRESTO (Grant No. JPMJPR2597), and RIKEN Incentive Research Project. We thank N. Nagaosa, K. Ishizaka, and D. Nakamura for fruitful discussions and experimental support.

**Author contributions**

Z.L, Y.To and Y.Ta. designed and coordinated this study. Z.L. grew the single crystals with the help of M.K.. Z.L. conducted the in-house single crystal X-ray diffraction, Laue diffraction, SEM and EDS measurements with the help of M.K., M.N., K.K. and A.K. S.K. and Z.L. conducted the single crystal synchrotron X-ray diffraction experiment at ambient pressure. S.K. and T.A. solved the crystal structure of ambient pressure. Z.L. and K.K. conducted the in-house single crystal X-ray diffraction at ambient pressure. Z.L. conducted the ambient and high-pressure electrical transport measurements with the help of M.N., M.K., C.T. and Y.F.. Z.L. and M.N. analyzed the high-pressure electrical anisotropy transport data using Montgomery method. X.L., K.M.Y, S.L., and X.Y. conducted the STEM and EDS measurements. K.S. and N. N. conducted TEM sample preparation. Z.L and Y.Ta. wrote the manuscript. All authors discussed the experimental results and contributed to the manuscript. Z.L and M.N. contributed equally to this work.

## Competing interests

The authors declare no competing interests.

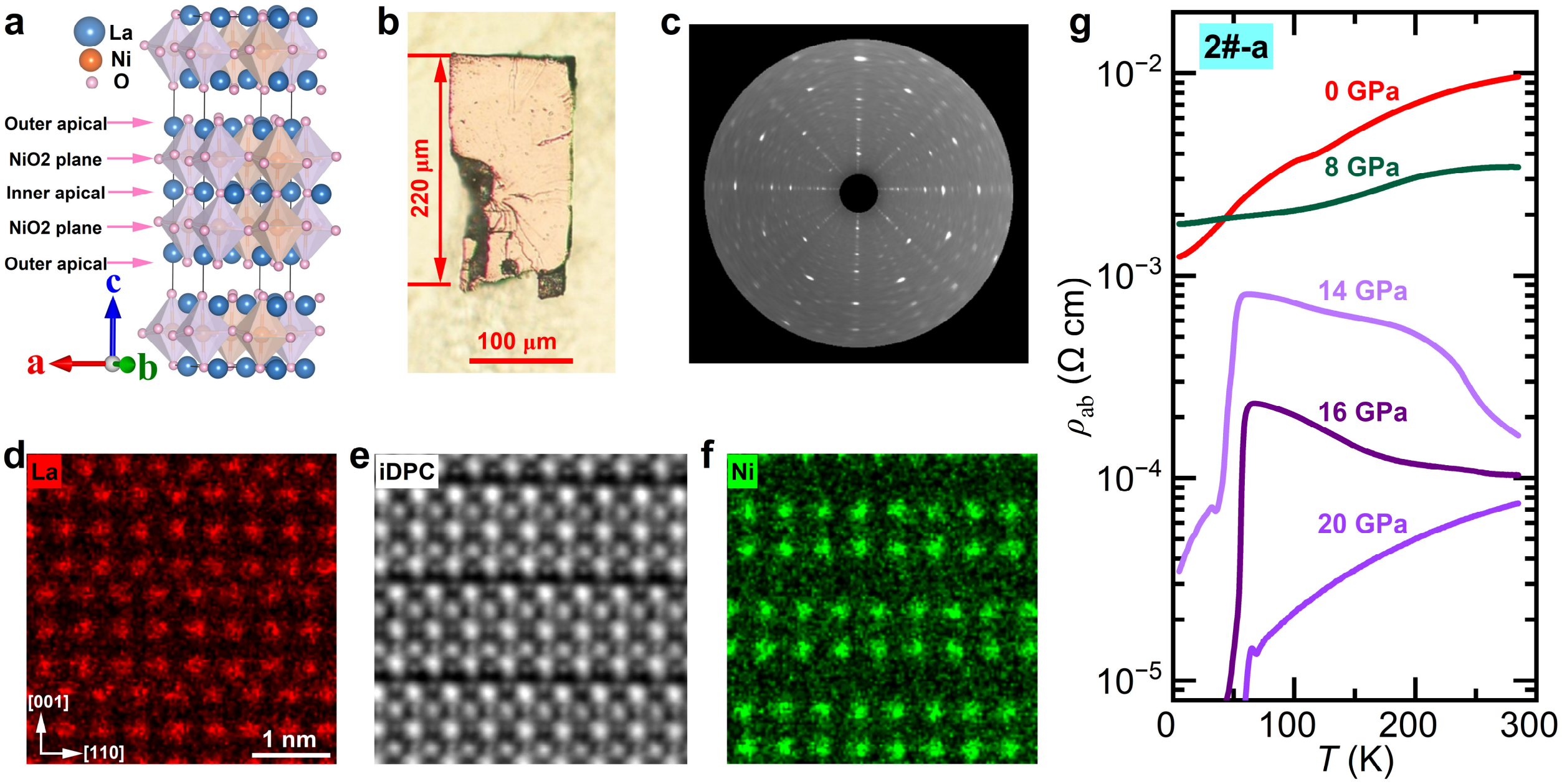

**Fig. 1| Structure and basic properties of $La_3Ni_2O_7$ single crystals. a**, Schematic crystal structure of $La_3Ni_2O_7$ with space group *A*m2m. Light yellow and purple colors indicate the different $NiO_2$ octahedra due to charge ordering at ambient temperature and ambient pressure. **b**, Polarization microscopy topography, **c**, Laue diffraction pattern for an *ab*-plane cleavage surface, and **e**, high-resolution integrated differential phase contrast (iDPC) images of single crystals, indicating $NiO_2$ bilayer structure without any intergrowth of other Ruddlesden-Popper phases. EDS maps from STEM for (**d**) La and (**f**) Ni, respectively. **g**, Temperature dependence of the in-plane resistivity of sample 2#-a at selected pressures plotted on a logarithmic scale.

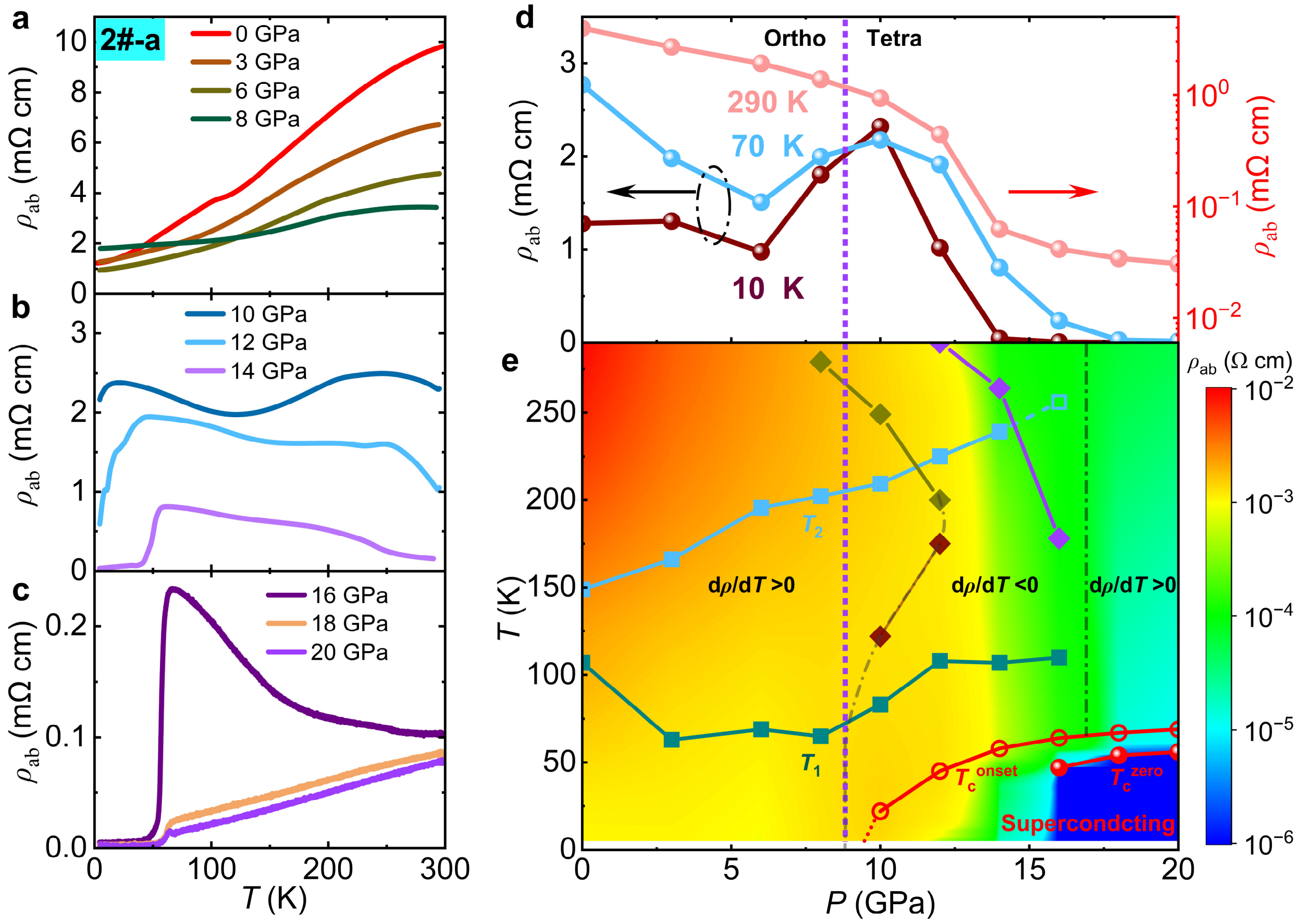

**Fig. 2| Pressure-induced superconductivity of $La_3Ni_2O_7$ single crystal 2#-a. a-c**, Temperature dependence of the in-plane resistivity under various hydrostatic pressures up to 20 GPa plotted on a linear scale. **d**, Pressure dependence of the resistivity at selected temperatures. **e**, *P*-*T* phase diagram of single crystal 2#-a with a logarithmic resistivity scale. In panel (**e**), $T_1$, $T_2$, $T_c^{onset}$, and $T_c^{zero}$ denote the characteristic temperatures of the density wave, spin-density wave, superconducting onset, and zero-resistance state, respectively. Dark yellow, wine, and violet diamonds indicate resistivity maximum, resistivity minimum, and onset of the sharp resistivity upturn, respectively. The dark-grey dash-dotted lines indicate the approximate boundary between regions with positive and negative temperature coefficients of the resistivity ($d\rho/dT > 0$ and $d\rho/dT < 0$, respectively). The purple dashed line in (**d,e**) indicates the approximate boundary between the orthorhombic (Ortho) and tetragonal (Tetra) structural phases.

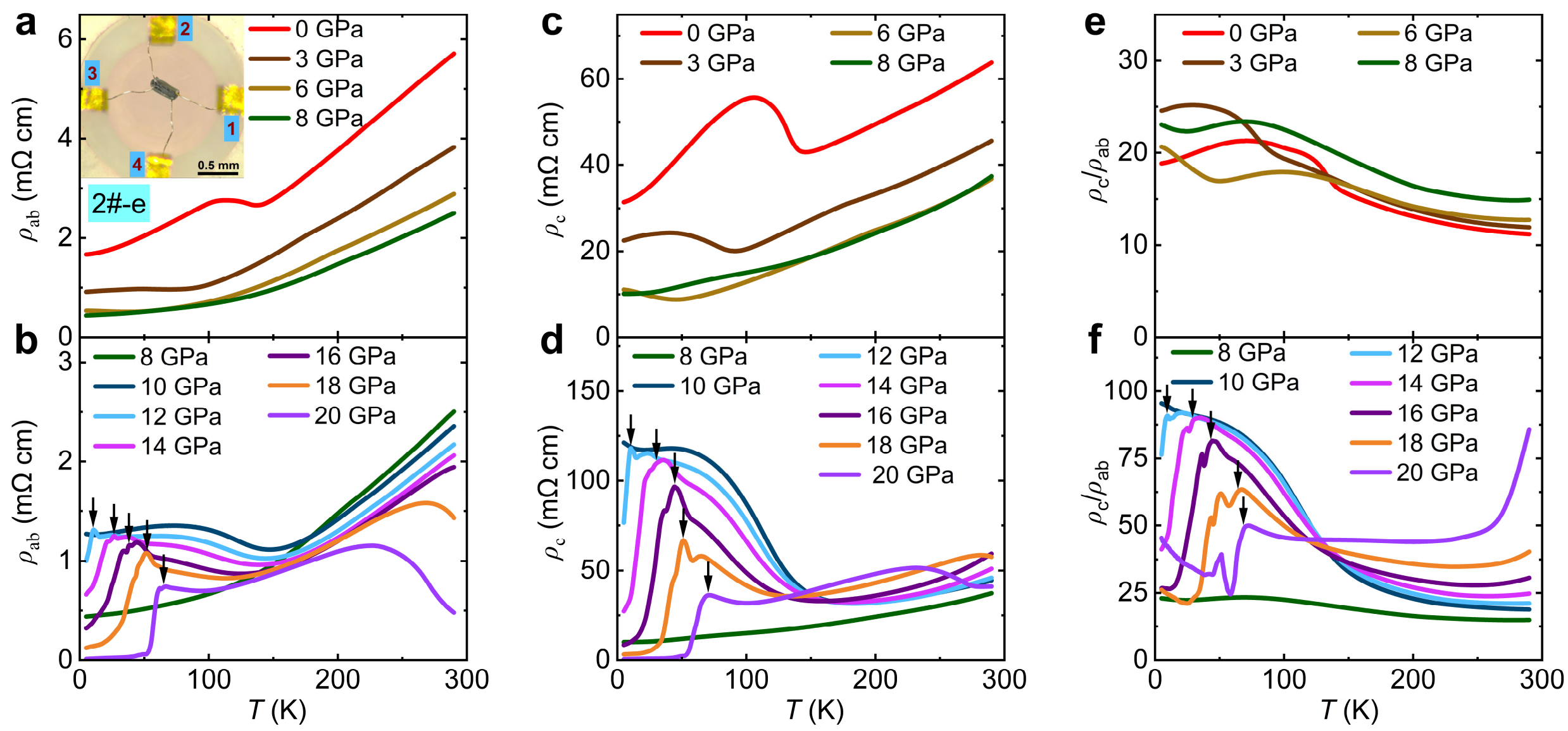

**Fig. 3| Electrical transport anisotropy of $La_3Ni_2O_7$ single crystal 2#-e.** Temperature dependence of (a, b) resistivity within the *ab*-plane, (c, d) resistivity along the *c*-axis, and (e, f) anisotropy, under various hydrostatic pressures up to 20 GPa measured on sample 2#-e. The inset of (a) shows a photo of single crystal 2#-e with gold electrodes inside the gasket. Scale bar indicates a length of 0.5 mm. The arrows in (b, d, f) indicate the $T_c^{onset}$ of superconductivity.

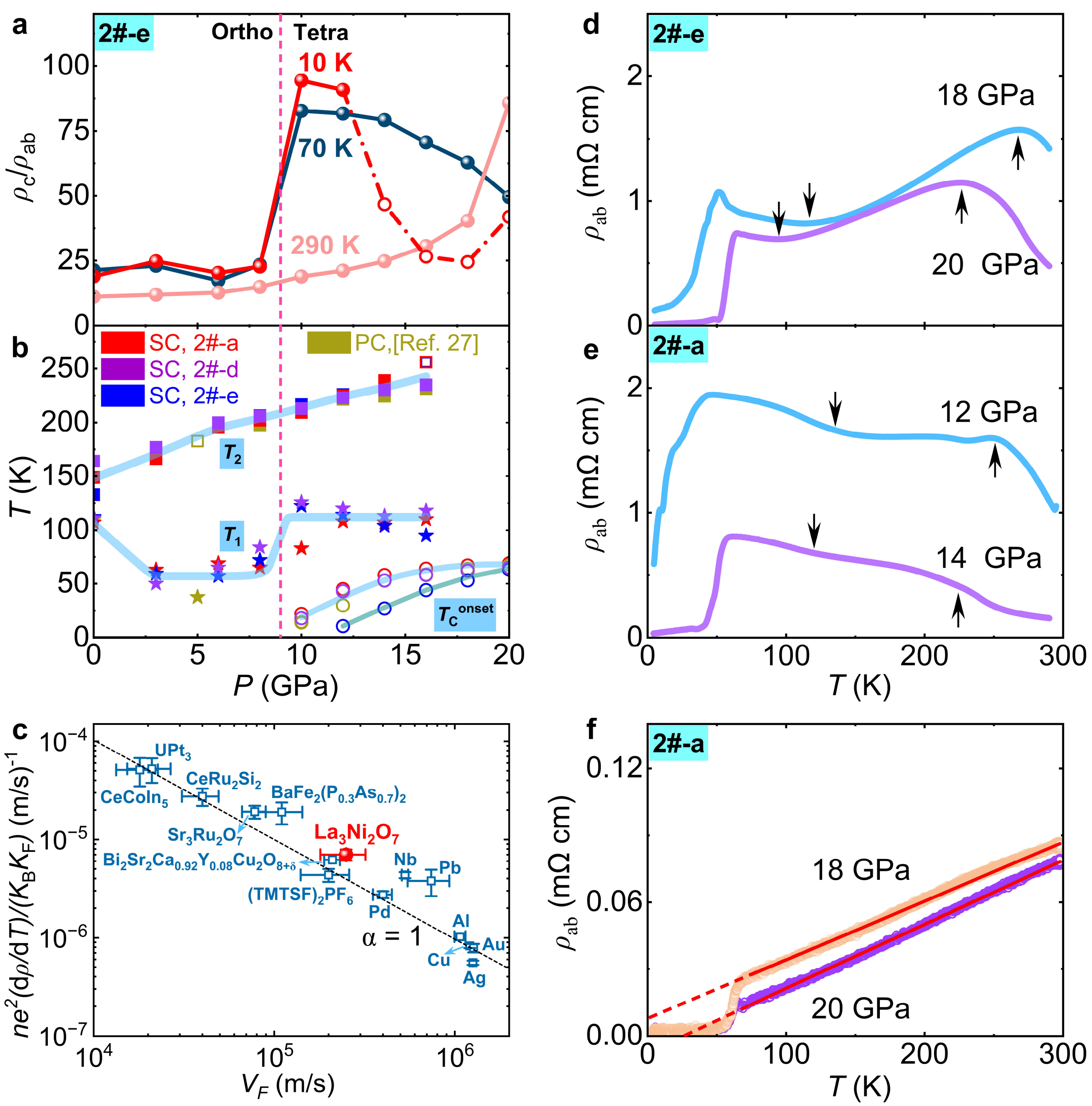

**Fig. 4| Electrical transport properties and Planckian dissipation in $La_3Ni_2O_7$ single crystals.** **a**, Pressure dependence of the anisotropy at selected temperatures for sample 2#-e. **b**, Pressure dependence of the characteristic temperatures for different single-crystals (SC) and a polycrystalline (PC) sample (the latter is reproduced from Ref. [27]). The assignments of the characteristic temperatures are shown in **Extended Data Fig. 1** (single crystal 2#-a), **Extended Data Figs. 2,3** (single crystal 2#-e) and **Extended Data Figs. 6,7** (single crystal 2#-d). The pink

dashed line in (**a,b**) indicates the approximate boundary between the orthorhombic (Ortho) and tetragonal (Tetra) structural phases. Light-blue and dark-cyan solid lines in panel **b** are guides to the eye. **c**, Relation between normalized slope of the $T$-linear resistivity and the Fermi velocity across a wide range of correlated and conventional metals. The dashed black line represents the Planckian dissipation limit ($\alpha = 1$) of the scattering rate $1/\tau \sim \alpha k_B T/\hbar$. The red symbol marks $La_3Ni_2O_7$, which falls close to the universal Planckian scattering observed in diverse quantum materials. Other data plotted with blue symbols are reproduced from Ref. [28]. **d**, **e**, **f**, Temperature dependence of the *ab*-plane resistivity for (**d)** single crystal 2#-e at 18 GPa and 20 GPa, for (**e**) single crystal 2#-a at 12 GPa and 14 GPa, and (**f**) for single crystal 2#-a at 18 GPa and 20 GPa. The arrows in (**d,e**) indicate the corresponding high-$T$ crossover. In the normal state above $T_c$, both datasets exhibit a $T$-linear dependence over a broad temperature range. The dashed red lines represent linear fits to the normal-state resistivity, which yield small slopes of 0.264 $\mu\Omega$ cm $K^{-1}$ and 0.287 $\mu\Omega$ cm $K^{-1}$, together with near-zero residual resistivity intercepts at 18 and 20 GPa, respectively.

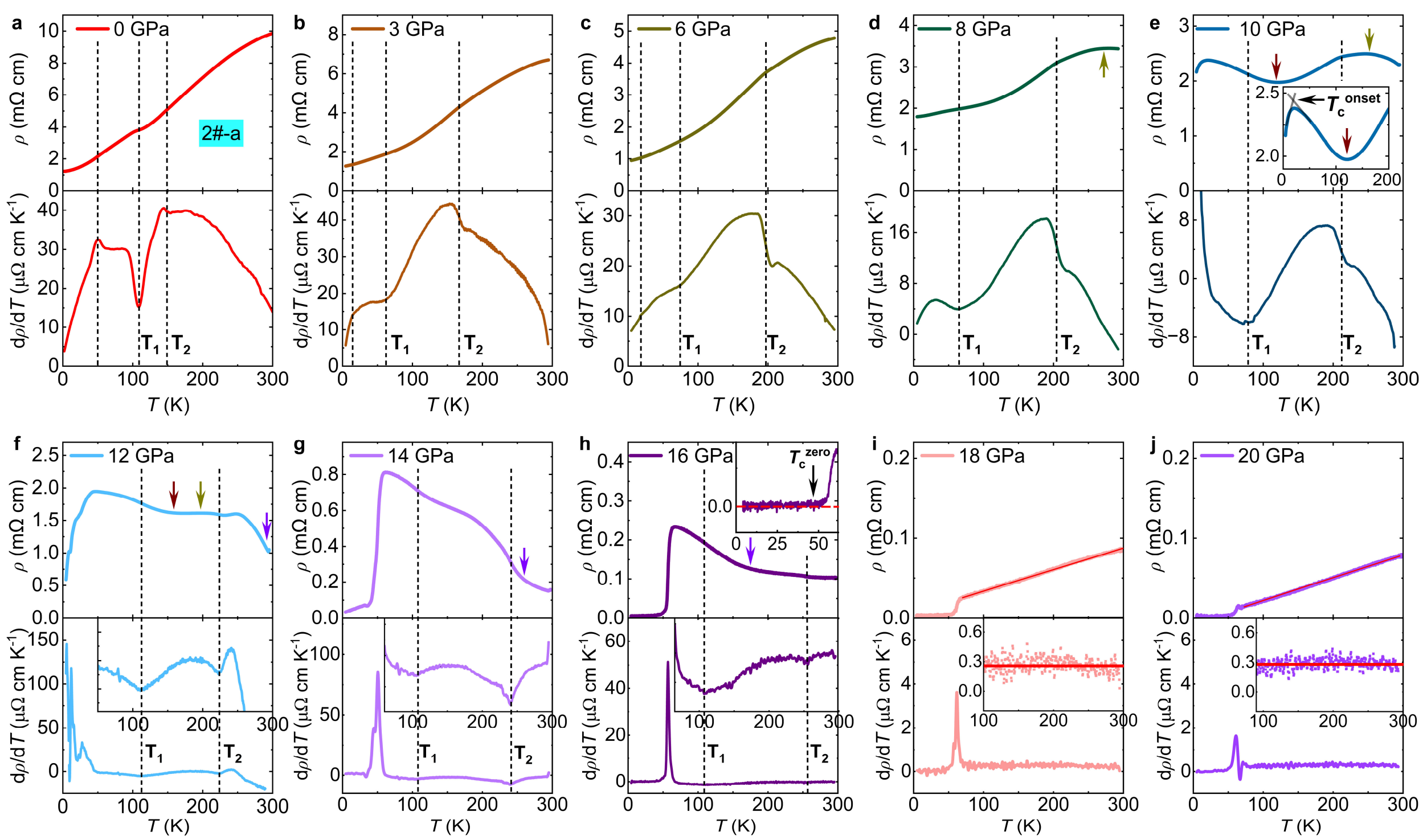

**Extended Data Fig. 1| Determination of characteristic temperatures from the pressure-dependent in-plane resistivity and its derivative of $La_3Ni_2O_7$ single crystal 2#-a. a-j**, Temperature dependence of the in-plane resistivity (upper panel) and its derivative (lower panel) under various pressures. The dashed lines in panels (**a-h**) indicate the definitions of the characteristic temperatures. $T_c^{onset}$ and $T_c^{zero}$ in panel (**e**) and panel (**h**) represent the superconducting onset temperature and the temperature corresponding to zero resistivity of the superconducting transition, respectively. Dark yellow, wine, violet arrows in (**d-h**) indicate the resistivity maximum, the resistivity minimum, and the onset of the sharp resistivity upturn, respectively. The red lines in the insets of lower panels (i) and (j) indicate the fitted slope of the $T$-linear resistivity at 18 GPa and 20 GPa, respectively.

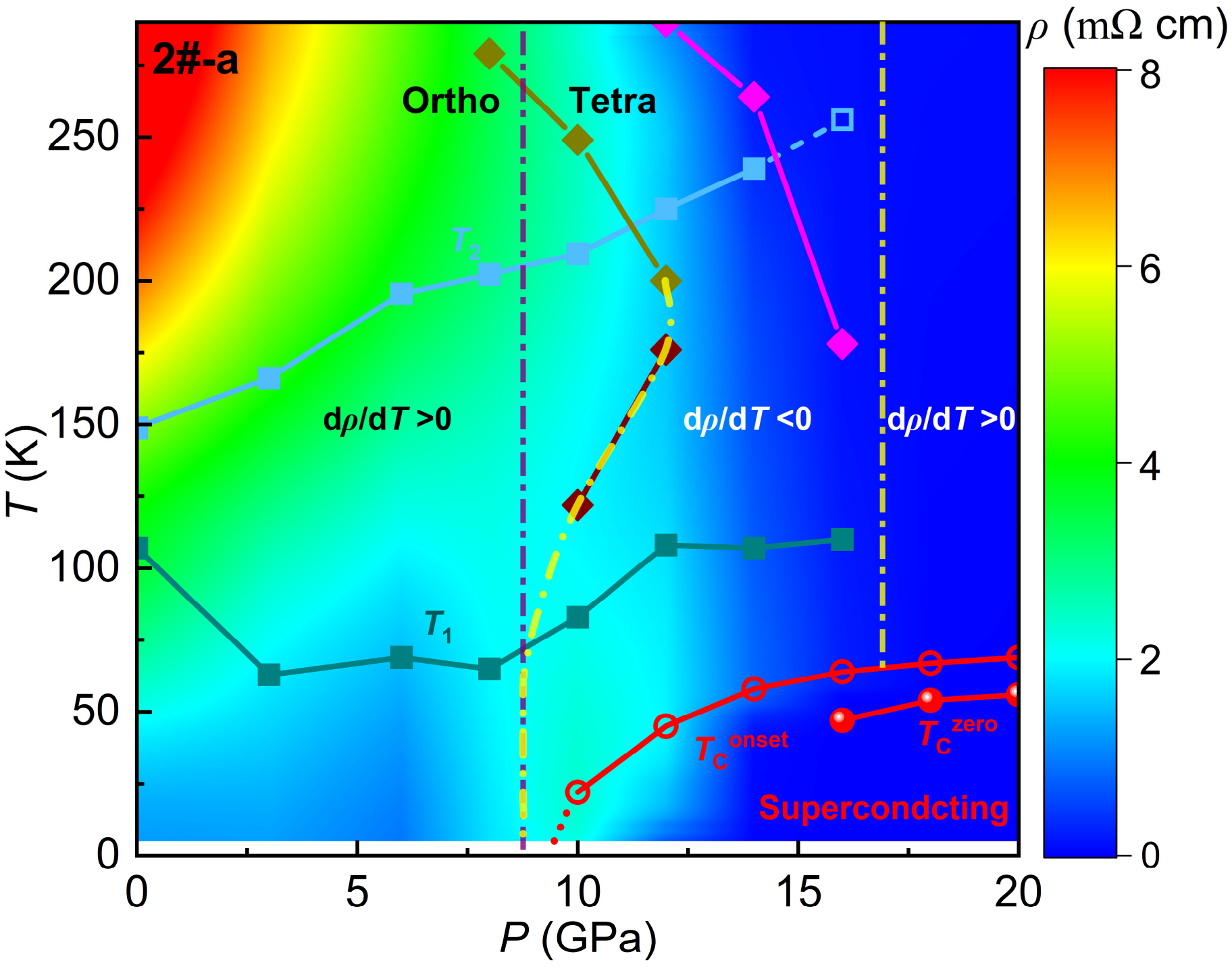

**Extended Data Fig. 2| *P-T* phase diagram of $La_3Ni_2O_7$ single crystal 2#-a with a linear resistivity scale.** $T_1$, $T_2$, $T_c^{onset}$, and $T_c^{zero}$ denote the characteristic temperatures of the density wave, spin-density wave, superconducting onset, and zero-resistance state, respectively. Dark yellow, wine, and pink diamonds indicate resistivity maximum, resistivity minimum, and onset of the sharp resistivity upturn, respectively. The yellow dash-dotted lines indicate the approximate boundary between regions with positive and negative temperature coefficients of the resistivity ($d\rho/dT > 0$ and $d\rho/dT < 0$, respectively). The purple dash-dotted line indicates the approximate boundary between the orthorhombic (Ortho) and tetragonal (Tetra) structural phases.

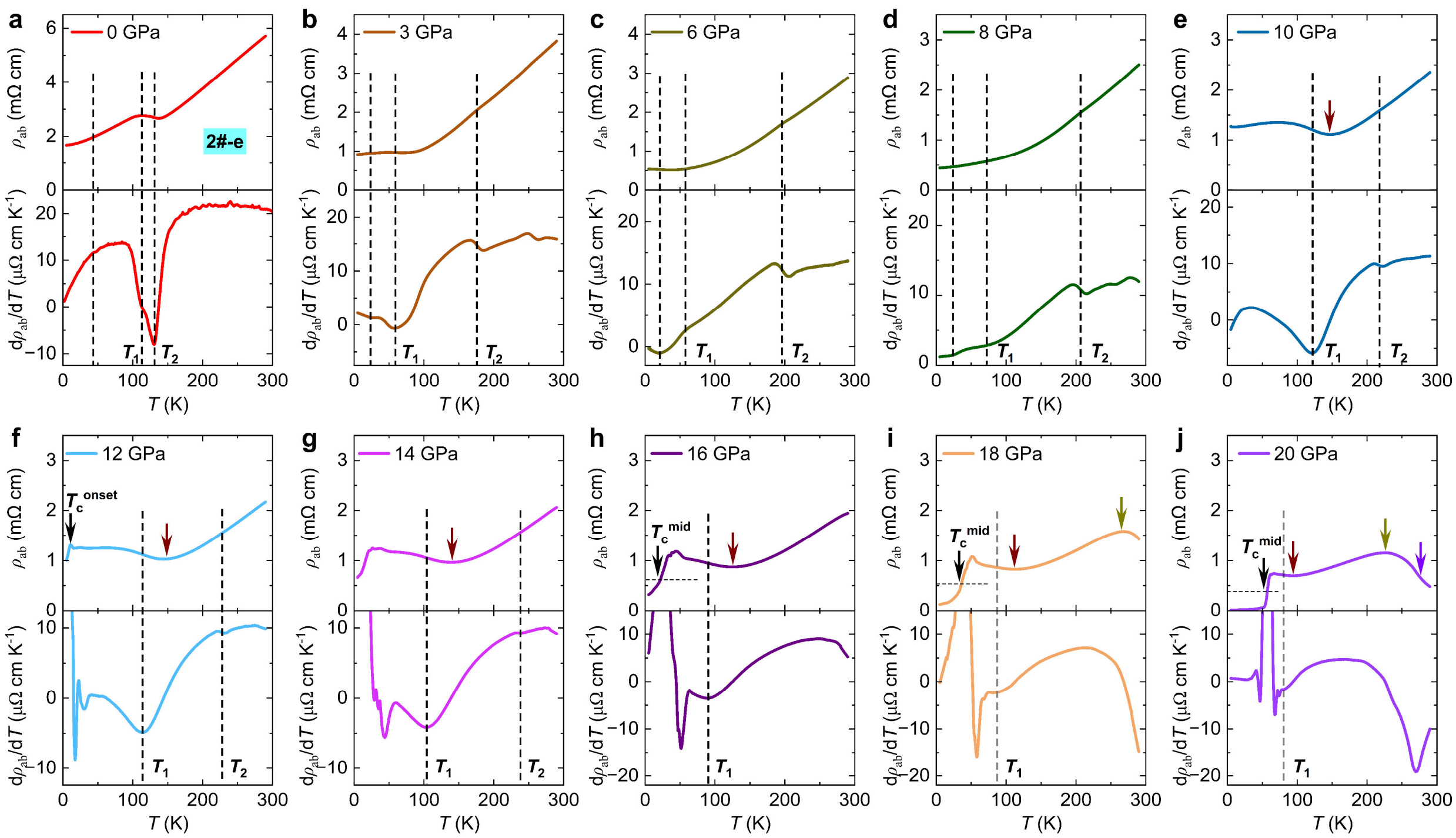

**Extended Data Fig. 3| Determination of characteristic temperatures from the pressure-dependent in-plane resistivity and its derivative of $La_3Ni_2O_7$ single crystal 2#-e. a-j**, Temperature dependence of the in-plane resistivity (upper panel) and its derivative (lower panel) under various pressures. The dashed lines indicate the definitions of the characteristic temperatures. Dark yellow, wine, violet arrows in panels (**e-j**) indicate the resistivity maximum, the resistivity minimum, and the onset of the sharp resistivity upturn, respectively. $T_c^{onset}$ and $T_c^{mid}$ in panels (**f, h-j**) represent the superconducting onset temperature and the temperature corresponding to the midpoint (50% drop in resistivity) of the superconducting transition, respectively.

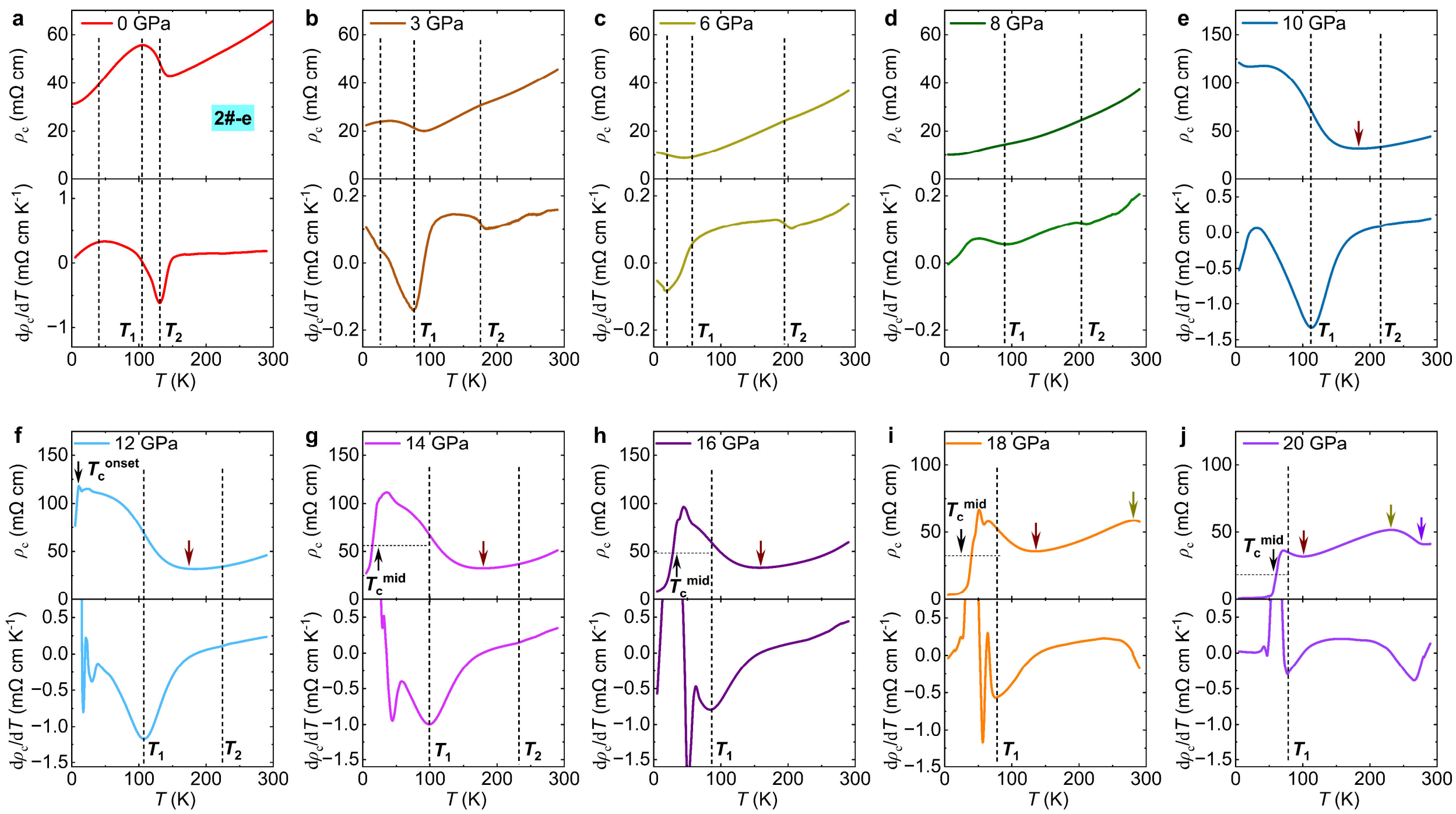

**Extended Data Fig. 4| Determination of characteristic temperatures from the pressure-dependent out-of-plane resistivity and its derivative of $La_3Ni_2O_7$ single crystal 2#-e. a-j**, Temperature dependence of the out-of-plane resistivity (upper panel) and its derivative (lower panel) of single crystal 2#-e under various pressures. The dashed lines indicate the definitions of the characteristic temperatures. Dark yellow, wine, and violet arrows in panels (**e-j**) indicate the resistivity maximum, the resistivity minimum, and the onset of the sharp resistivity upturn, respectively. $T_c^{onset}$ and $T_c^{mid}$ in panels (**f, g-j**) represent the superconducting onset temperature and the temperature corresponding to the midpoint (50% drop in resistivity) of the superconducting transition, respectively.

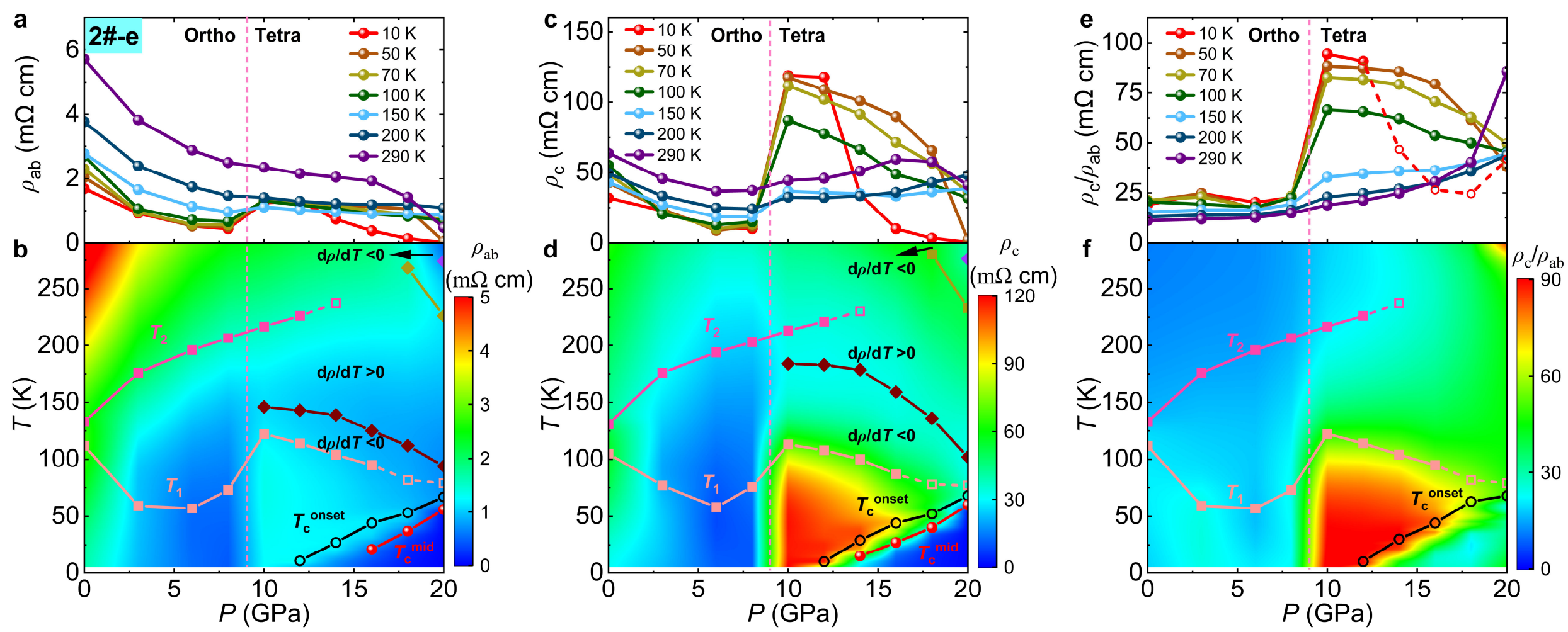

**Extended Data Fig. 5| Pressure evolution of anisotropic electrical transport and the corresponding *P-T* phase diagrams of $La_3Ni_2O_7$ single crystal 2#-e. a**,**c**,**e**, Pressure dependences of the in-plane resistivity, the out-of-plane resistivity, and the resistivity anisotropy, respectively, measured at selected temperatures for single crystal 2#-e. **b**,**d**,**f**, Corresponding *P*-*T* phase diagrams constructed from the transport data. Dark yellow, wine, violet diamonds in panels **(b, d)** indicate the characteristic temperatures of the resistivity maximum and the resistivity minimum (onset of the sharp resistivity upturn), respectively. $T_c^{onset}$ and $T_c^{mid}$ represent the superconducting onset temperature and the temperature corresponding to the midpoint (50% drop in resistivity) of the superconducting transition, respectively. The characteristic temperatures used to construct the *P-T* phase diagrams are determined from the in-plane resistivity (or its temperature derivative) for **(b)**, the out-of-plane resistivity (or its temperature derivative) for **(d)**, and the resistivity anisotropy (or the temperature derivative of the in-plane resistivity) for **(f)**. The pink dashed lines in (**a-f**) indicate the approximate boundary between the orthorhombic (Ortho) and tetragonal (Tetra) structural phases.

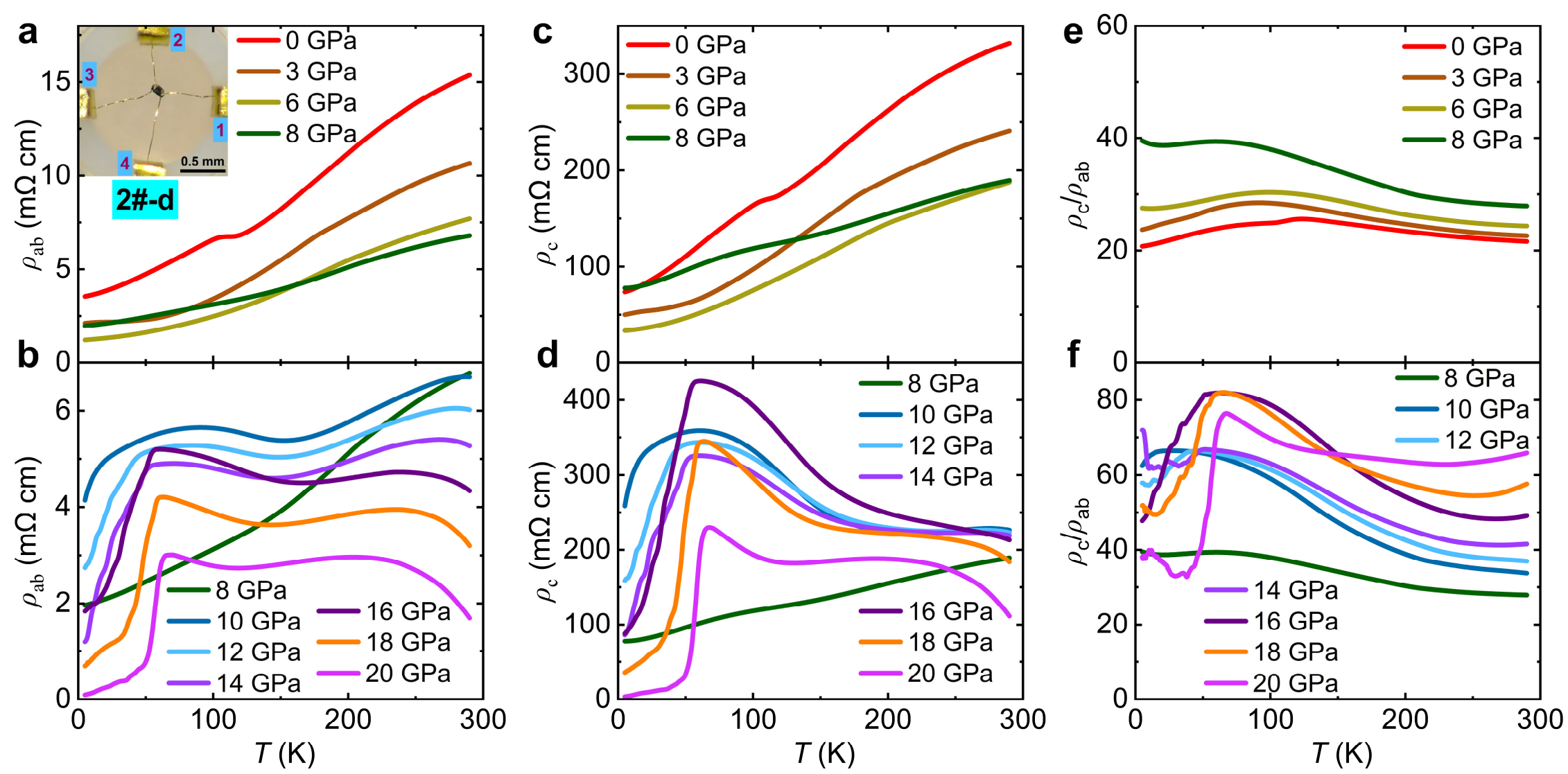

**Extended Data Fig. 6| Electrical transport anisotropy of $La_3Ni_2O_7$ single crystal 2#-d.** Temperature dependence of (**a, b**) the resistivity within the *ab*-plane, (**c, d**) the resistivity along the *c*-axis, and (**e, f**) the anisotropy, under various hydrostatic pressures up to 20 GPa measured on sample 2#-d. The inset of (**a**) shows the topography of single crystal 2#-d with gold electrodes inside the gasket. The scale bar indicates a length of 0.5 mm.

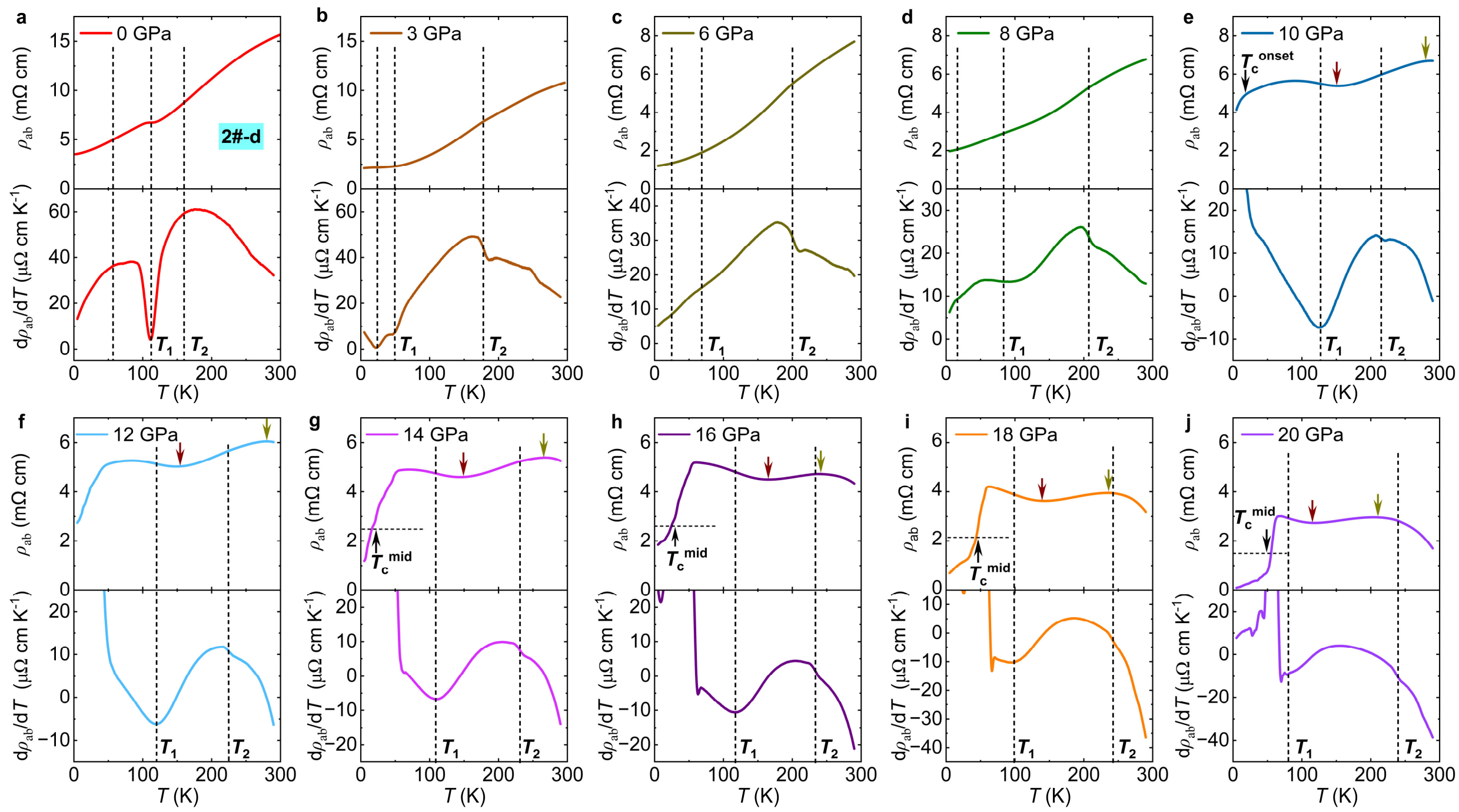

**Extended Data Fig. 7| Determination of characteristic temperatures from the pressure-dependent in-plane resistivity and its derivative of $La_3Ni_2O_7$ single crystal 2#-d. a-j**, Temperature dependence of the in-plane resistivity (upper panel) and its derivative (lower panel) under various pressures. The dashed lines indicate the definitions of the characteristic temperatures. Dark yellow and wine arrows in panels (**e-j**) indicate the resistivity maximum, the resistivity minimum, and the onset of the sharp resistivity upturn, respectively. $T_c^{onset}$ and $T_c^{mid}$ in panels (**f, g-j**) represent the superconducting onset temperature and the temperature corresponding to the midpoint (50% drop in resistivity) of the superconducting transition, respectively.

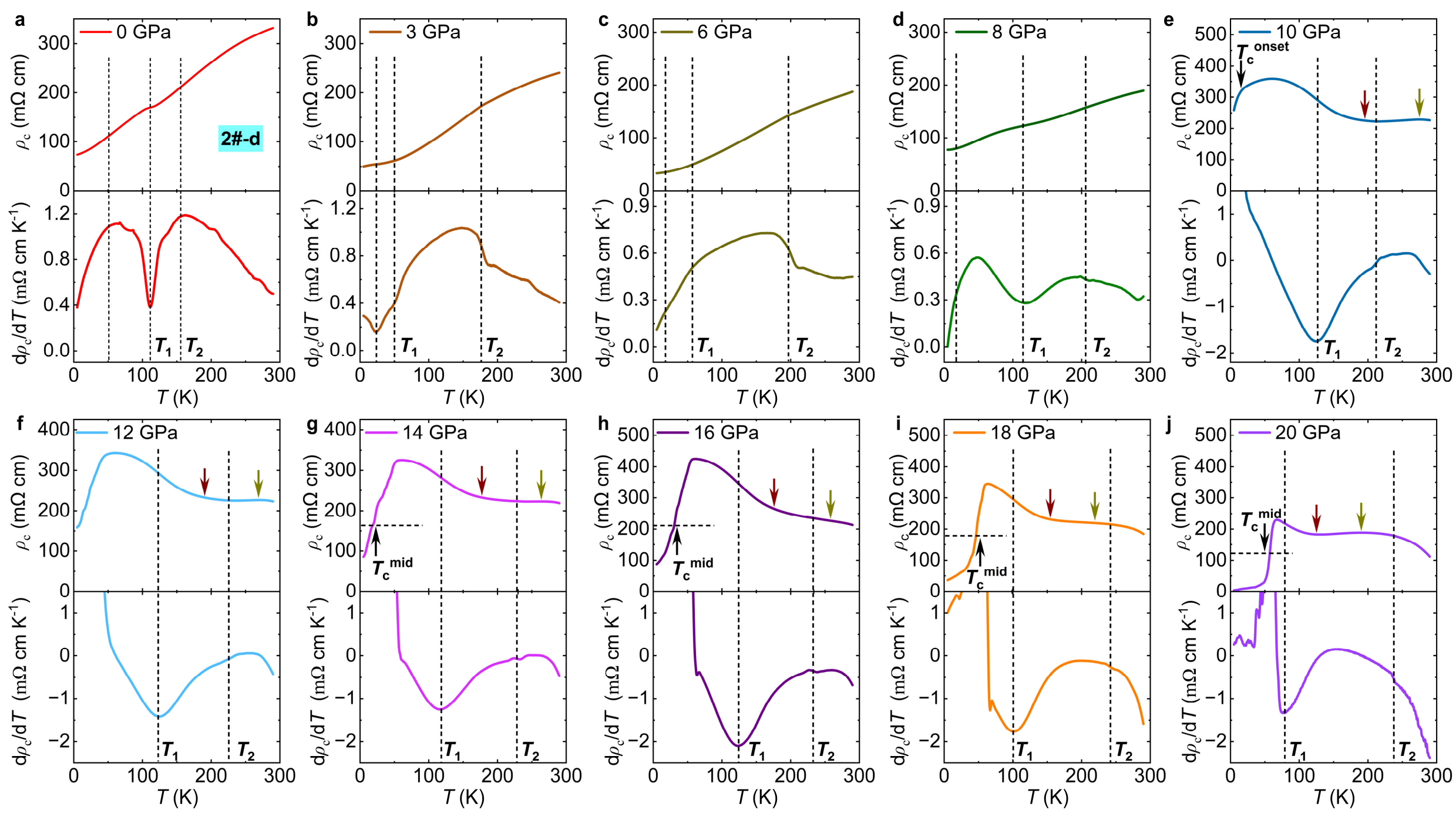

**Extended Data Fig. 8| Determination of characteristic temperatures from the pressure-dependent out-of-plane resistivity and its derivative of $La_3Ni_2O_7$ single crystal 2#-d. a-j**, Temperature dependence of the out-of-plane resistivity (upper panel) and its derivative (lower panel) under various pressures. The dashed lines indicate the definitions of the characteristic temperatures. Dark yellow and wine arrows in panels (**e-j**) indicate the resistivity maximum and the resistivity minimum (onset of the sharp resistivity upturn), respectively. $T_c^{onset}$ and $T_c^{mid}$ in panels (**f, g-j**) represent the superconducting onset temperature and the temperature corresponding to the midpoint (50% drop in resistivity) of the superconducting transition, respectively.

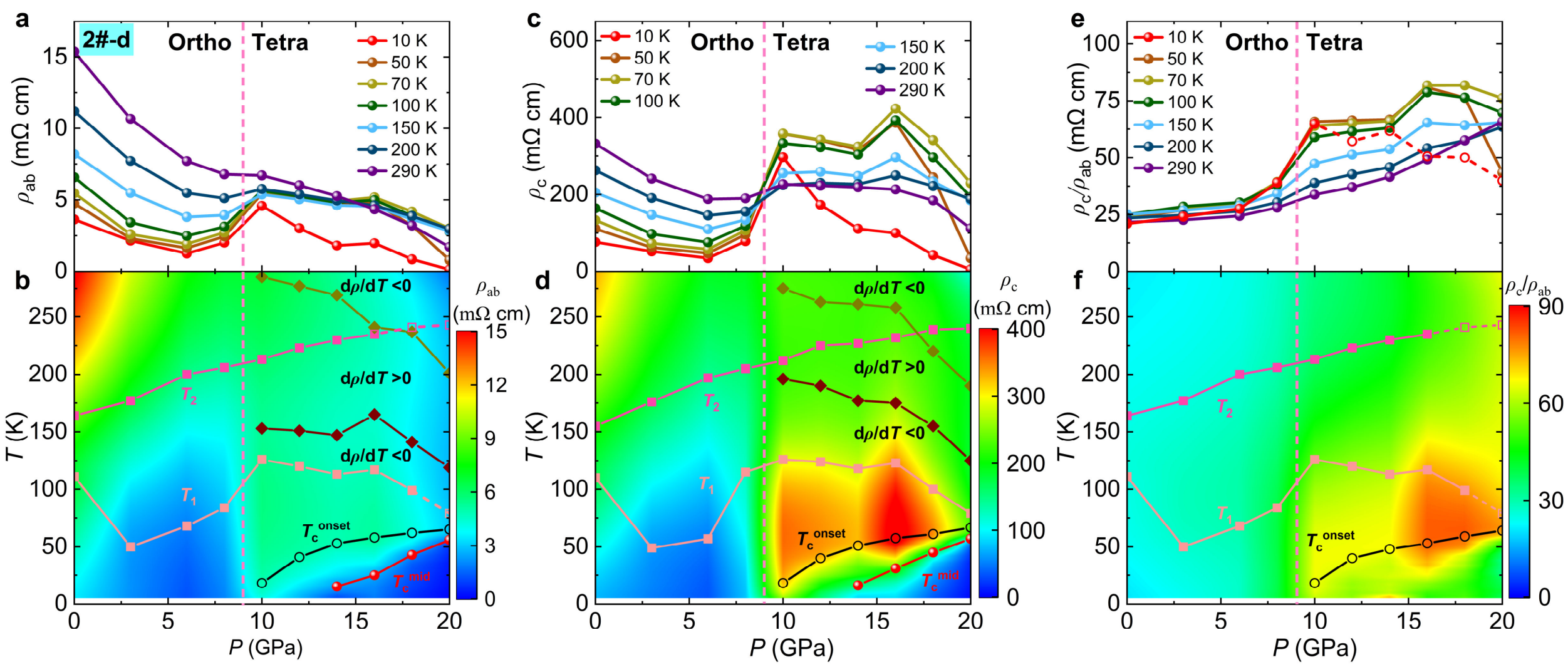

**Extended Data Fig. 9| Pressure evolution of the anisotropic electrical transport and the corresponding *P-T* phase diagrams of $La_3Ni_2O_7$ single crystal 2#-d. a**,**c**,**e**, Pressure dependences of the in-plane resistivity, out-of-plane resistivity, and resistivity anisotropy, respectively, measured at selected temperatures on single crystal 2#-d. **b**,**d**,**f**, Corresponding *P*-*T* phase diagrams constructed from the transport data. Dark yellow and wine diamonds indicate the characteristic temperature of the resistivity maximum and the resistivity minimum (onset of the sharp resistivity upturn), respectively. $T_c^{onset}$ and $T_c^{mid}$ represent the superconducting onset temperature and the temperature corresponding to the midpoint (50% drop in resistivity) of the superconducting transition, respectively. The characteristic temperatures used to construct the *P*-*T* phase diagrams are determined from the in-plane resistivity (or its temperature derivative) for **(b)**, the out-of-plane resistivity (or its temperature derivative) for **(d)**, and the resistivity anisotropy (or the temperature derivative of the in-plane resistivity) for **(f)**. The pink dashed lines in (**a-f**) indicate the approximate boundary between the orthorhombic (Ortho) and tetragonal (Tetra) structural phases.

**Supplementary Information for**

# Density-wave phases, anisotropic transport, and Planckian dissipation in single crystals of the superconductor $La_3Ni_2O_7$

Zhehong Liu[1,2†*], Masamichi Nakajima[1†], Markus Kriener[1], Shunsuke Kitou[3], Xiaowei Lyu[1], Chieko Terakura[1], Kosuke Karube[1], Ka Man Yip[4], Sorin Lazar[4], Nobuto Nakanishi[5], Keiko Shimada[1], Akiko Kikkawa[1], Yukako Fujishiro[1,6,7,8], Xiuzhen Yu[1], Taka-hisa Arima[1,3], Yoshinori Tokura[1,7,9*], Yasujiro Taguchi[1,2*]

[1]*RIKEN Center for Emergent Matter Science* (*CEMS*)*, Wako 351-0198, Japan*

[2]*RIKEN Baton Zone Program, TRIP Headquarters, Wako 351-0198, Japan*

[3]*Department of Advanced Materials Science, The University of Tokyo, Kashiwa 277-8561, Japan*

[4]*Materials and Structural Analysis Division, Thermo Fisher Scientific, Achtseweg Noord 5, Eindhoven, Netherlands*

[5]*Materials and Structural Analysis Division, Thermo Fisher Scientific, FEI Company Japan Ltd. Tokyo140-0002, Japan*

[6]*RIKEN Pioneering Research Institute* (*PRI*), *Wako 351-0198, Japan*

[7]*Department of Applied Physics, The University of Tokyo, Tokyo 113-8656, Japan*

[8]*Institute of Engineering Innovation, The University of Tokyo, Tokyo 113-8656, Japan*

[9]*Tokyo College, The University of Tokyo, Tokyo 113-8654, Japan*

**Crystal structure at ambient pressure**

X-ray diffraction measurements on a 50 × 50 × 20 μm$^3$ crystal at 300 K and ambient pressure reveals diffraction patterns consistent with an orthorhombic lattice (**Supplementary Table S1** and **Supplementary Fig. S4**). Although additional reflections originating from minor domains are observed, their intensities are less than 1% of those of the main domain and are therefore neglected in the following analysis. The diffraction data satisfy the *A*-centering reflection condition (*hkl*: *k*+*l* = 2*n*). Although many experimental studies on the structure of $La_3Ni_2O_7$ have reported the *Amam* structure at ambient pressure [1,2,3], inspection of the *h*0*l* plane reveals the presence of reflections with *h* = 2*n*+1, indicating a violation of the extinction condition associated with the *a*-glide plane perpendicular to the *b*-axis. These observations unambiguously indicate that the space group is *Am*2*m*.

Structural refinement and analysis (as shown in **Supplementary Fig. S5**) based on the main-domain intensities using the *Am*2*m* model yield a structure consistent with the recent report [4], featuring two distinct Ni sites in the unit cell (**Fig. 1a** and **Supplementary Table S2**). The refinement reveals rotations of the $NiO_6$ octahedra about the *a*-axis (Glazer notation: $a^-a^-c^0$ [5]) together with clear differences in the $NiO_6$ octahedral volume between the two Ni sites, consistent with charge order. Refinement of the oxygen site occupancies further indicates no detectable oxygen deficiency within experimental accuracy (**Supplementary Table S3**).

**Estimation of the scattering rate and Planckian coefficient $\alpha$ in $La_3Ni_2O_7$**

We estimate the dimensionless parameter *a* that appears in the formula of the Planckian scattering rate $\frac{1}{\tau} = \alpha \frac{k_B T}{\hbar}$, as follows.

As shown in the Supplementary Information of Ref. [6],

$$\alpha = A \frac{e^2}{2\pi k_{\mathrm{B}} d} \sum_i M_i k_{\mathrm{F}i} v_{\mathrm{F}i}$$

where *A* is the coefficient of the *T*-linear resistivity, *e* the electron charge, *d* the bilayer-bilayer spacing of 10.26 Å, and $M_i$, $k_{\mathrm{F}i}$, and $v_{\mathrm{F}i}$ are the multiplicity, average Fermi wave number, and average Fermi velocity of Fermi surface $i$, respectively. Here, in the case of $La_3Ni_2O_7$, two Fermi surfaces (called a and b surfaces) are observed by an angle-resolved photoemission spectroscopy measurement [7]. For both surfaces, $M_i = 1$. From the experimentally determined Fermi surfaces and band dispersions, we evaluate

$k_{\mathrm{F\alpha}} = 0.29 \pm 0.02\ [1/\text{Å}]$, $k_{\mathrm{F\beta}} = 0.77 \pm 0.02\ [1/\text{Å}]$, $v_{\mathrm{F\alpha}} = (2.6 \pm 0.3) \times 10^5$ [m/s], $v_{\mathrm{F\beta}} = (1.2 \pm 0.3) \times 10^5$ [m/s].

According to first-principles band calculations [8,9], electron transfer interactions are enhanced by ca. 20% under the application of 20 GPa. Therefore, we use the values of $v_{\mathrm{Fa}}$ and $v_{\mathrm{Fb}}$ enhanced by 20% from the ambient pressure values. On the basis of these values, we estimate $\alpha = 1.6 \pm 0.4$.

As for the definition of the parameters $n$, $k_{\mathrm{F}}$, and $v_{\mathrm{F}}$ adopted in Fig.4c, we follow the convention presented in the Supplementary Information of Ref. [6]. For the present two-dimensional system $La_3Ni_2O_7$ with two Fermi surfaces, $n$, $k_{\mathrm{F}}$, and $v_{\mathrm{F}}$ are defined as

$$n = n_\alpha + n_\beta$$

$$k_{\mathrm{F}}^2 = \sum_i k_{\mathrm{F}i}^2$$

$$v_{\mathrm{F}} = \left(\sum_i k_{\mathrm{F}i} v_{\mathrm{F}i}\right) \Big/ k_{\mathrm{F}}$$

and are calculated to be $(1.1\pm0.1)\times10^{28}$ [m$^{-3}$], $0.82 \pm 0.03$ [A$^{-1}$], and $(2.5\pm0.7)\times10^5$ [m s$^{-1}$], respectively. Here, the $v_{\mathrm{F}}$ value is also multiplied by 1.2 due to the anticipated enhancement of the transfer interaction under 20 GPa. By using these values, we obtain

$$\frac{ne^2}{k_{\mathrm{B}} k_{\mathrm{F}}} \frac{d\rho}{dT} = (7.0 \pm 0.9) \times 10^{-6}\ \ [(\mathrm{m/s})^{-1}]$$

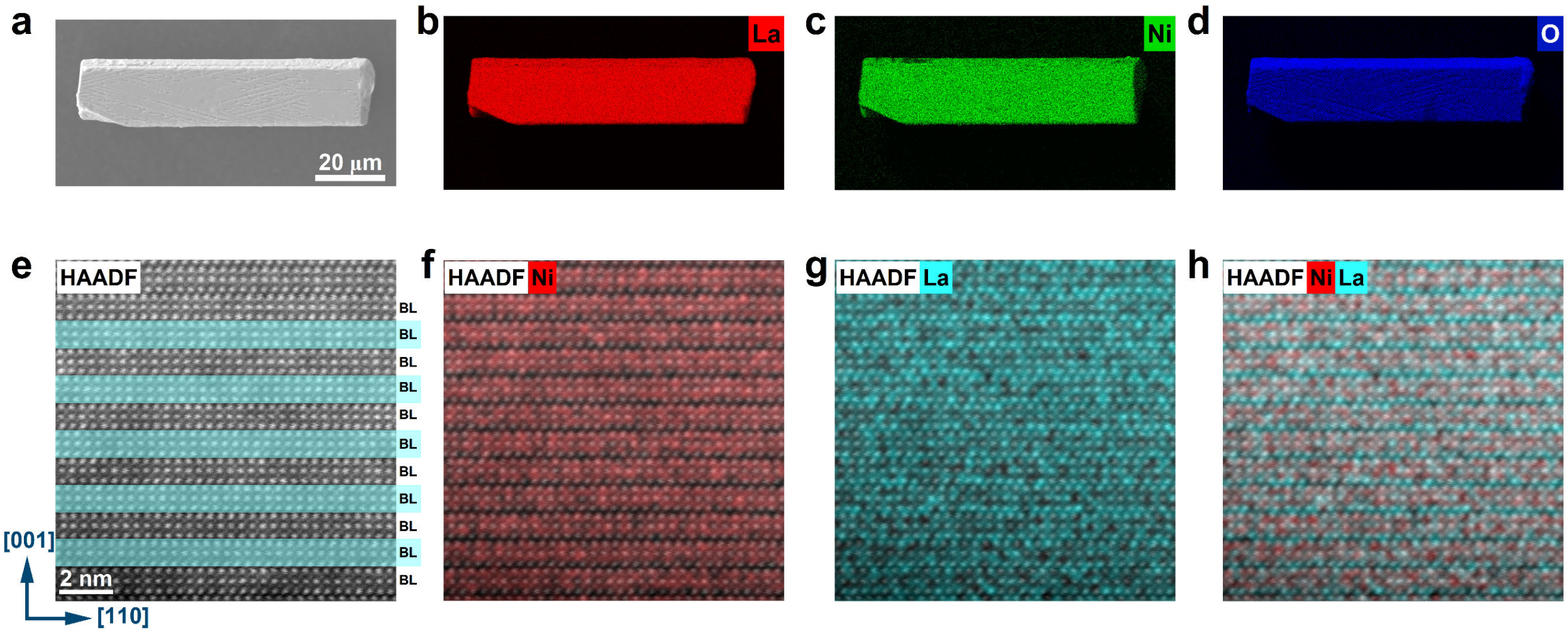

**Supplementary Fig. S1| a-d,** SEM image and EDS mapping of a typical $La_3Ni_2O_7$ single crystal. **e-h,** STEM image and EDS maps from STEM for La, and Ni. BL stands for bilayer.

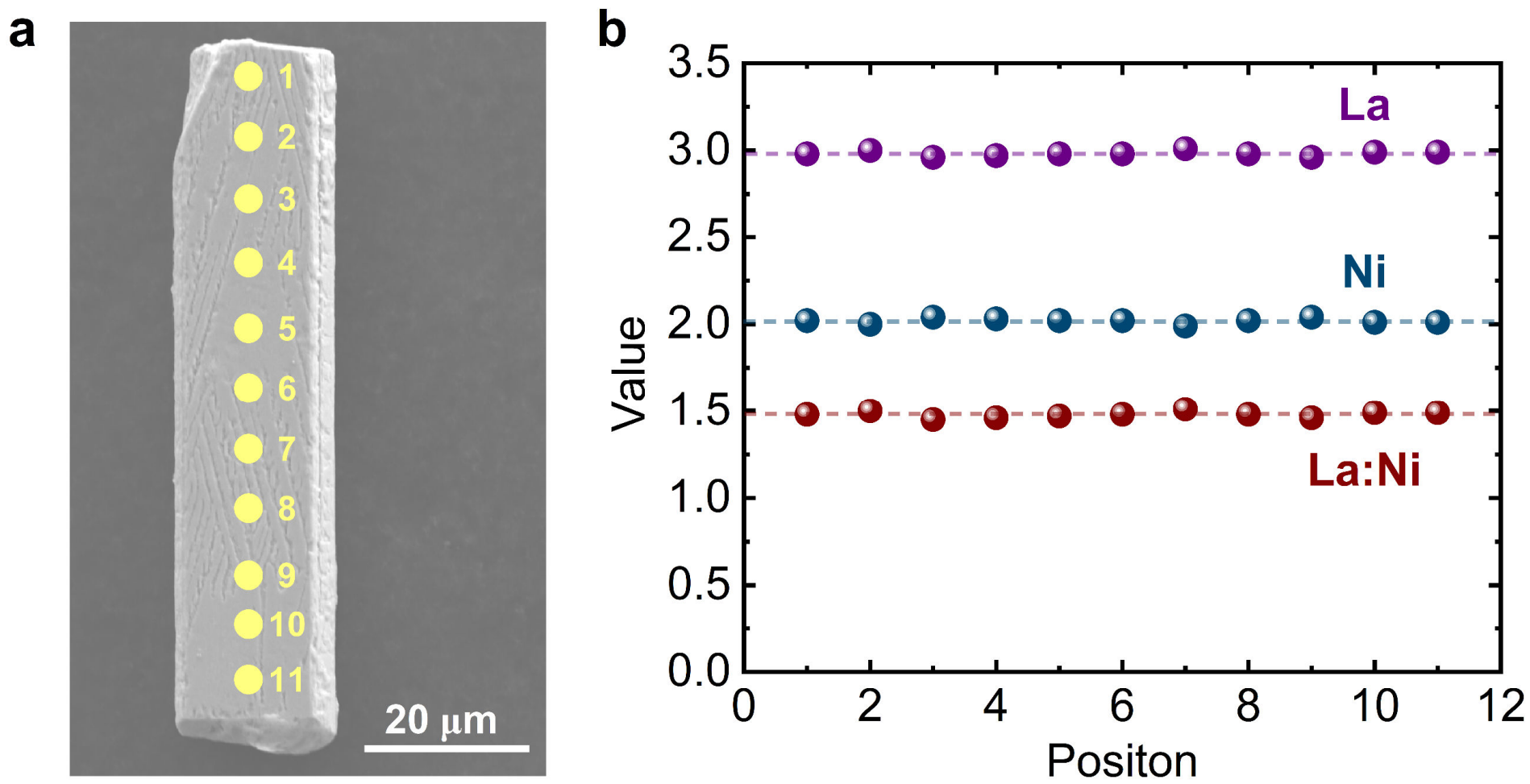

**Supplementary Fig. S2| a**, SEM image of a representative $La_3Ni_2O_7$ single crystal showing the eleven positions selected for quantitative EDS analysis. **b**, Quantitative EDS analysis of the La content, Ni content and La:Ni ratio measured at each position. The dashed lines indicate the average values of 2.98(3), 2.02(3) and 1.48(3), respectively, confirming excellent compositional homogeneity and near-stoichiometric composition throughout the crystal.

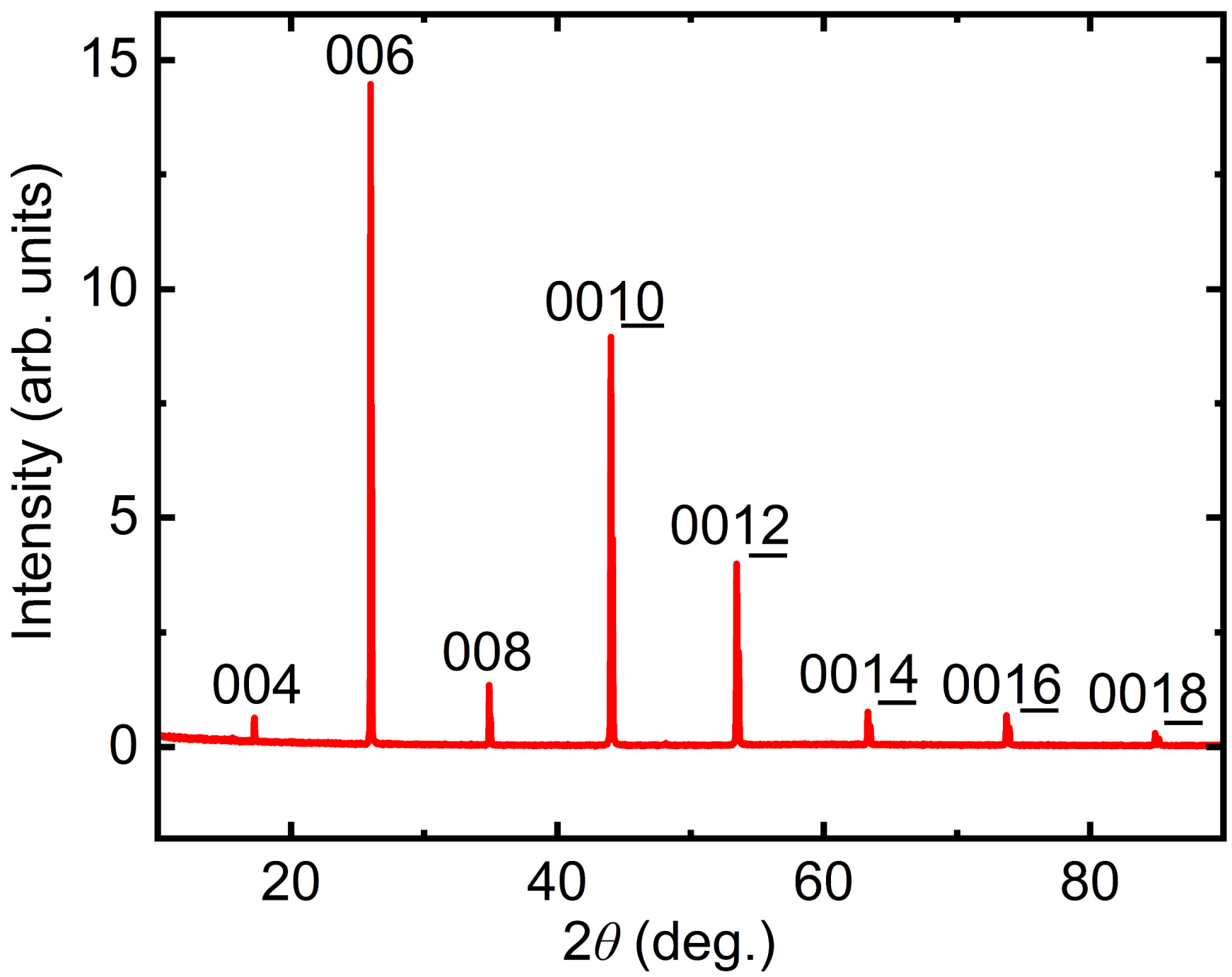

**Supplementary Fig. S3|** In-house X-ray diffraction pattern of an *ab*-plane plate of a typical $La_3Ni_2O_7$ single crystal showing only (00*l*) reflections.

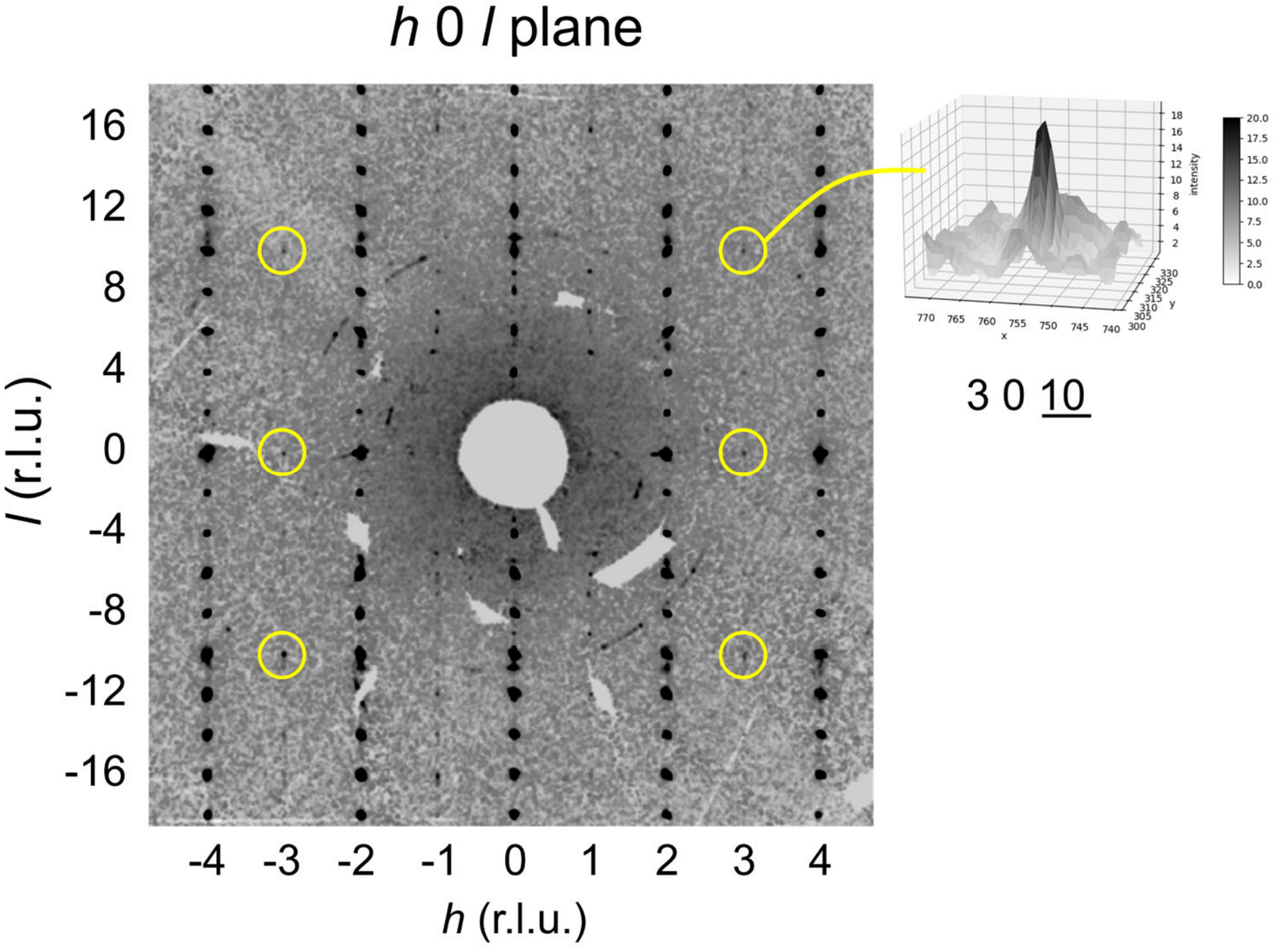

**Supplementary Fig. S4|** Synchrotron X-ray diffraction data on the *h*0*l* plane of $La_3Ni_2O_7$ at 300 K. The presence of *h*0*l* reflections with *h* = odd indicates the violation of the extinction condition associated with the *a*-glide plane perpendicular to the *b*-axis, confirming the *Am*2*m* space group.

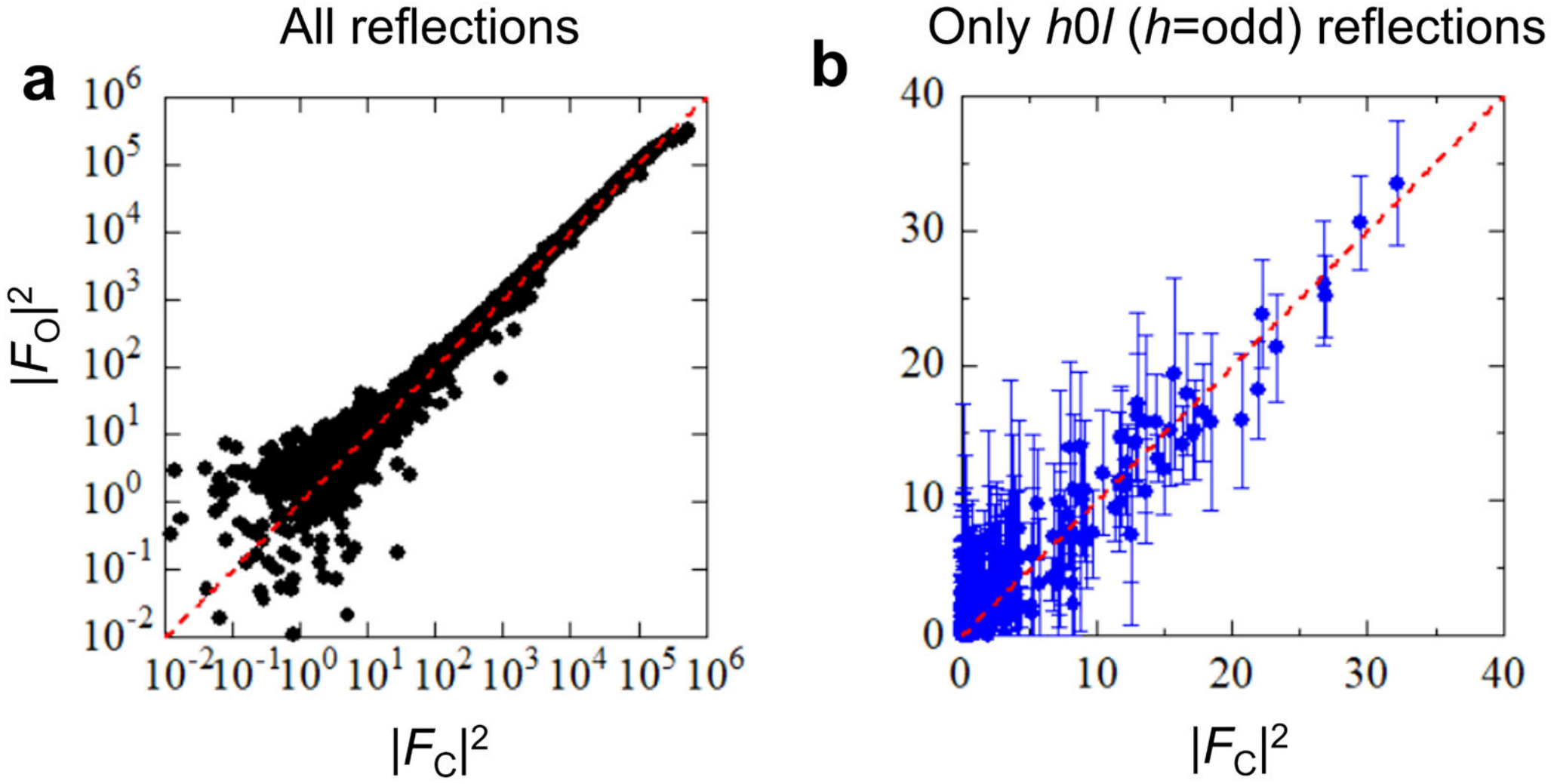

**Supplementary Fig. S5| a**, Logarithmic and **b**, linear $|F_O|^2$-$|F_C|^2$ plots obtained from the structural analysis of $La_3Ni_2O_7$ at 300 K, obtained without assuming O vacancies. Black and blue circles represent all reflections and *h*0*l* reflections with *h* = odd, respectively.

**Supplementary Table S1| Summary of the crystallographic data of $La_3Ni_2O_7$ at 300 K.**

| | |
|---|---|
| Wavelength (Å) | 0.34300 |
| Crystal dimension ($\mu m^3$) | 50 × 50 × 20 |
| Space group | *Am*2*m* |
| *a* (Å) | 5.39478(7) |
| *b* (Å) | 5.45148(7) |
| *c* (Å) | 20.5604(3) |
| *V* ($Å^3$) | 604.672(14) |
| Z | 4 |
| *F*(000) | 1132 |
| $(\sin\theta/\lambda)_{max}$ ($Å^{-1}$) | 1.62 |
| Number of total reflections | 41370 |
| Average redundancy | 3.868 |
| Completeness (%) | 96.64 |
| Number of unique reflections ($I > 3\sigma$ / all) | 8952/ 10696 |
| Without assuming O vacancies | |
| $N_{parameters}$ | 72 |
| $R_1$ ($I > 3\sigma$/ all) (%) | 2.80 / 3.04 |
| $wR_2$ ($I > 3\sigma$/ all) (%) | 3.21 / 3.28 |
| GOF ($I > 3\sigma$/ all) | 1.36 / 1.27 |
| Flack parameters (*a*,*b*,*c* : -*a*,-*b*,-*c*) | 0.39(7) : 0.61 |
| Assuming O vacancies | |
| $N_{parameters}$ | 78 |
| $R_1$ ($I > 3\sigma$/ all) (%) | 2.80 / 3.05 |
| $wR_2$ ($I > 3\sigma$/ all) (%) | 3.20 / 3.27 |
| GOF ($I > 3\sigma$/ all) | 1.36 / 1.27 |
| Flack parameters (*a*,*b*,*c* : -*a*,-*b*,-*c*) | 0.39(7) : 0.61 |

**Supplementary Table S2| Refined structural parameters of $La_3Ni_2O_7$ at 300 K obtained without assuming O vacancies.**

| Atom | Wyckoff position | Site symmetry | *x* | *y* | *z* | Occupancy |
|---|---|---|---|---|---|---|
| La1 | 4e | *m*.. | 1/2 | 0.494491(19) | 0.679588(4) | 1 |
| La2 | 2a | *mm*2 | 0 | 0.00000(2) | 1/2 | 1 |
| La3 | 4d | *m*.. | 0 | 0.011417(19) | 0.679789(4) | 1 |
| La4 | 2b | *mm*2 | 1/2 | -0.50152(2) | 1/2 | 1 |
| Ni1 | 4d | *m*.. | 0 | 0.50477(5) | 0.596138(11) | 1 |
| Ni2 | 4e | *m*.. | 1/2 | -0.00051(4) | 0.595762(10) | 1 |
| O1 | 8f | 1 | 0.2534(3) | 0.2469(3) | 0.58932(3) | 1 |
| O2 | 8f | 1 | 0.2538(3) | -0.2425(3) | 0.60524(3) | 1 |
| O3 | 4d | *m*.. | 0 | -0.0302(4) | 0.79674(5) | 1 |
| O4 | 2b | *mm*2 | 1/2 | -0.0438(5) | 1/2 | 1 |
| O5 | 2a | *mm*2 | 0 | 0.0466(5) | 0 | 1 |
| O6 | 4e | *m*.. | 1/2 | -0.4597(4) | 0.20563(6) | 1 |
| Atom | $U_{11}$ ($10^{-2}$ Å$^2$) | $U_{22}$ ($10^{-2}$ Å$^2$) | $U_{33}$ ($10^{-2}$ Å$^2$) | $U_{12}$ ($10^{-2}$ Å$^2$) | $U_{13}$ ($10^{-2}$ Å$^2$) | $U_{23}$ ($10^{-2}$ Å$^2$) |
| La1 | 0.590(3) | 0.558(5) | 0.383(2) | 0 | 0 | 0.0284(18) |
| La2 | 0.683(4) | 0.697(5) | 0.552(3) | 0 | 0 | 0 |
| La3 | 0.598(3) | 0.534(4) | 0.392(2) | 0 | 0 | -0.0155(17) |
| La4 | 0.716(4) | 0.723(5) | 0.544(3) | 0 | 0 | 0 |
| Ni1 | 0.365(4) | 0.347(7) | 0.473(5) | 0 | 0 | -0.017(4) |
| Ni2 | 0.324(4) | 0.304(6) | 0.447(4) | 0 | 0 | 0.021(4) |
| O1 | 0.747(15) | 0.800(18) | 1.147(18) | 0.232(12) | -0.03(2) | -0.01(2) |
| O2 | 0.723(14) | 0.698(15) | 1.347(18) | -0.266(11) | 0.02(3) | 0.04(3) |
| O3 | 1.17(4) | 1.05(4) | 0.54(3) | 0 | 0 | -0.02(3) |
| O4 | 1.25(6) | 1.02(6) | 0.42(4) | 0 | 0 | 0 |
| O5 | 1.52(7) | 1.04(7) | 0.43(4) | 0 | 0 | 0 |
| O6 | 1.60(5) | 1.29(5) | 0.62(3) | 0 | 0 | 0.18(4) |

**Supplementary Table S3| Refined structural parameters of $La_3Ni_2O_7$ at 300 K obtained with assuming O vacancies.**

| Atom | Wyckoff position | Site symmetry | *x* | *y* | *z* | Occupancy |
|---|---|---|---|---|---|---|
| La1 | 4e | *m*.. | 1/2 | 0.494472(19) | 0.679587(4) | 1 |
| La2 | 2a | *mm*2 | 0 | 0.00000(2) | 1/2 | 1 |
| La3 | 4d | *m*.. | 0 | 0.01140(2) | 0.679789(4) | 1 |
| La4 | 2b | *mm*2 | 1/2 | -0.50151(2) | 1/2 | 1 |
| Ni1 | 4d | *m*.. | 0 | 0.50477(5) | 0.596140(11) | 1 |
| Ni2 | 4e | *m*.. | 1/2 | -0.00052(5) | 0.595762(10) | 1 |
| O1 | 8f | 1 | 0.2533(3) | 0.2470(3) | 0.58936(3) | 1.020(4) |
| O2 | 8f | 1 | 0.2538(3) | -0.2424(3) | 0.60518(4) | 1.018(4) |
| O3 | 4d | *m*.. | 0 | -0.0302(4) | 0.79672(6) | 1.004(6) |
| O4 | 2b | *mm*2 | 1/2 | -0.0435(5) | 1/2 | 1.010(8) |
| O5 | 2a | *mm*2 | 0 | 0.0466(5) | 0 | 1.021(8) |
| O6 | 4e | *m*.. | 1/2 | -0.4599(4) | 0.20561(6) | 1.013(7) |
| Atom | $U_{11}$ ($10^{-2}$ Å$^2$) | $U_{22}$ ($10^{-2}$ Å$^2$) | $U_{33}$ ($10^{-2}$ Å$^2$) | $U_{12}$ ($10^{-2}$ Å$^2$) | $U_{13}$ ($10^{-2}$ Å$^2$) | $U_{23}$ ($10^{-2}$ Å$^2$) |
| La1 | 0.590(3) | 0.559(5) | 0.382(2) | 0 | 0 | 0.0288(18) |
| La2 | 0.683(4) | 0.698(5) | 0.552(3) | 0 | 0 | 0 |
| La3 | 0.597(3) | 0.532(5) | 0.391(2) | 0 | 0 | -0.0150(17) |
| La4 | 0.715(4) | 0.721(5) | 0.544(3) | 0 | 0 | 0 |
| Ni1 | 0.365(4) | 0.345(7) | 0.473(5) | 0 | 0 | -0.017(4) |
| Ni2 | 0.323(4) | 0.305(7) | 0.446(5) | 0 | 0 | 0.021(4) |
| O1 | 0.770(16) | 0.823(18) | 1.175(19) | 0.231(12) | -0.03(2) | -0.01(2) |
| O2 | 0.742(15) | 0.712(15) | 1.38(2) | -0.269(11) | 0.02(3) | 0.03(3) |
| O3 | 1.17(4) | 1.05(4) | 0.54(3) | 0 | 0 | -0.02(3) |
| O4 | 1.25(6) | 1.07(7) | 0.42(4) | 0 | 0 | 0 |
| O5 | 1.56(8) | 1.06(7) | 0.45(4) | 0 | 0 | 0 |
| O6 | 1.63(6) | 1.32(5) | 0.64(3) | 0 | 0 | 0.19(4) |

**Supplementary Table S4| Summary of the three $La_3Ni_2O_7$ single crystals (2#-a, d, e).**

| Sample number | Dimensions (μm$^3$) | $T_c^{onset}$ at 20 GPa (K) |
|---|---|---|
| 2#-a | 420 × 390 × 50 | 68 |
| 2#-d | 108 × 106 × 29 | 65 |
| 2#-e | 260 × 134× 63 | 67 |